\documentclass[a4paper,12pt]{elsarticle}

\usepackage{hyperref}
\usepackage{caption}
\usepackage{subcaption}
\journal{Journal of \LaTeX\ Templates}

\begin{document}

\begin{frontmatter}

\title{The Muon-Induced Neutron Indirect Detection EXperiment, MINIDEX}

\author[a]{I.~Abt}
\author[a]{A.~Caldwell}
\author[a]{C.~Carissimo}
\author[a]{C.~Gooch}
\author[a]{R.~Knei{\ss}l}
\author[a]{J.~Langford}
\author[a]{X.~Liu}
\author[a]{B.~Majorovits}
\author[a]{M.~Palermo\corref{cor1}}
\ead{palermo@mpp.mpg.de}
\author[a]{O.~Schulz}
\author[a]{L.~Vanhoefer}

\address[a]{Max-Planck-Institut f\"{u}r Physik, Munich, Germany}

\cortext[cor1]{Corresponding Author. Tel: +49-89-32354295}

\begin{abstract}
A new experiment to quantitatively 
measure neutrons induced by cosmic-ray muons in selected high-Z
materials is introduced.  
The design of the Muon-Induced Neutron Indirect Detection EXperiment, 
MINIDEX, and the results from its first data taking period 
are presented as well as future plans.
Neutron production in high-Z materials is
of particular interest as such materials are used for shielding in
low-background experiments.
The design of next-generation large-scale experiments searching for
neutrinoless double beta decay or direct interactions of dark matter 
requires reliable Monte Carlo simulations of background induced
by muon interactions. 
The first five months of operation already provided
a valuable data set on 
neutron production and neutron transport in lead.
A first round of comparisons between MINIDEX data 
and Monte Carlo predictions obtained with two GEANT4-based
packages is presented. 
The rate of muon-induced events is overall a factor three
to four higher in data 
than predicted by the Monte Carlo packages. 
In addition, the time evolution of the muon-induced signal is not 
well described by the simulations.

\end{abstract}


\end{frontmatter}


\section{Introduction}
\label{intro}
Muon-induced neutrons are not only an interesting 
physics topic by themselves, 
but they are also a potentially very important source of background in 
searches for new rare phenomena like 
neutrinoless double beta decay or directly observable
interactions of dark matter.

The  
Muon-Induced Neutron Indirect Detection EXperiment (MINIDEX)
was designed to provide data on the interactions of 
cosmic high-energy muons in high-Z materials.  
Previous measurements are not really 
consistent~\cite{allkofer,aglietta89,aglietta99,chen95,boehm,rapp,langford}.
Neither are the corresponding 
calculations~\cite{delorme,ryazhskaya,barton,allkofer,perkins}.  
One of the problems is, that for all measurements, it is hard to disentangle
the primary neutron production rate from the influence of
neutron transport. 
In addition, experiments have different energy thresholds 
for the detection of neutrons.
MINIDEX detects thermalised neutrons and thus, 
has no neutron-energy threshold.
Another problem is the determination of the background.
MINIDEX measures its own background to high precision 
simultaneously with the signal.

Neutrons produced by the interactions of high-energy muons with nuclei
are emitted in all directions and have energies of up to several~GeV.
As their energy exceeds the maximum energy of neutrons produced 
by natural radioactivity by several orders of magnitude,
they can be important for low-background experiments, 
even though they are much more 
rare~\cite{mei06,Araujo05,formaggio,wulandari-cresst,empl-fluka}.
The importance of muon-induced background 
will increase for the next generation
of low-background experiments, because the background level 
will have to be reduced by at least another order of magnitude 
compared to currently running 
experiments~\cite{kamlandZen,exo200,gerda-results,gerda-phase2}.

Traditionally, high-Z materials are used to shield low-background
experiments. Examples are the MAJORANA~\cite{majorana} and 
CDEX\,I~\cite{cdex} projects.
However, in such a configuration, muons produce high-energy neutrons 
in the shield, close to the active volume.
Therefore, besides going deeper underground,
techniques to veto muons are commonly used.
However, moderated neutrons can 
create unstable states close to or in the active volume
of an experiment which decay too slowly to be vetoed against. 
A lot of effort went into 
Monte Carlo packages~\cite{mei06,cdms2,geant3,geant4,fluka} 
to facilitate the evaluation of
such backgrounds. Nevertheless, all Monte Carlo predictions 
depend on the input assumptions about
the primary neutron-production processes, neutron transport
and neutron thermalisation.
As a result, the predictions vary substantially for the different 
packages~\cite{musun,mei06,Araujo05,empl-fluka,wang,mei14,reichhart,marino,lindote,wulandari-cresst}; 
summaries and discussions in~\cite{horn,kluck,palermo}.  

The data accumulated with MINIDEX provide
information on muon-induced neutron production
and transport by detecting thermalised neutrons.
For its first runs, MINIDEX was operated using lead as the target material. 
However, it was designed to provide data sets 
for a number of selected high-Z  materials.
The first data, taken from July 15 to November 25, 2015, 
are presented as well a comparison with GEANT-based
Monte Carlo predictions.

\section{The experimental setup}
\label{setup}

MINIDEX was designed such that it can be operated remotely. 
It is compact to fit into relatively small facilities, can
be moved easily and the target material can be exchanged. 
It  has to be operated underground to suppress background from cosmogenic
neutrons.     
As the experiment is relatively small, shallow sites are required
to have a large enough muon flux to obtain reasonable data rates.
MINIDEX was installed for its first data taking period, run\,I, in the 
T\"ubingen Shallow Underground Laboratory (TSUL) in July, 2015.
The nominal overburden of the TSUL corresponds to $\approx$\,16\,meter water 
equivalent~\cite{TSUL}.

\begin{figure}[!h]
  \begin{center}
    \includegraphics[scale = 0.35]{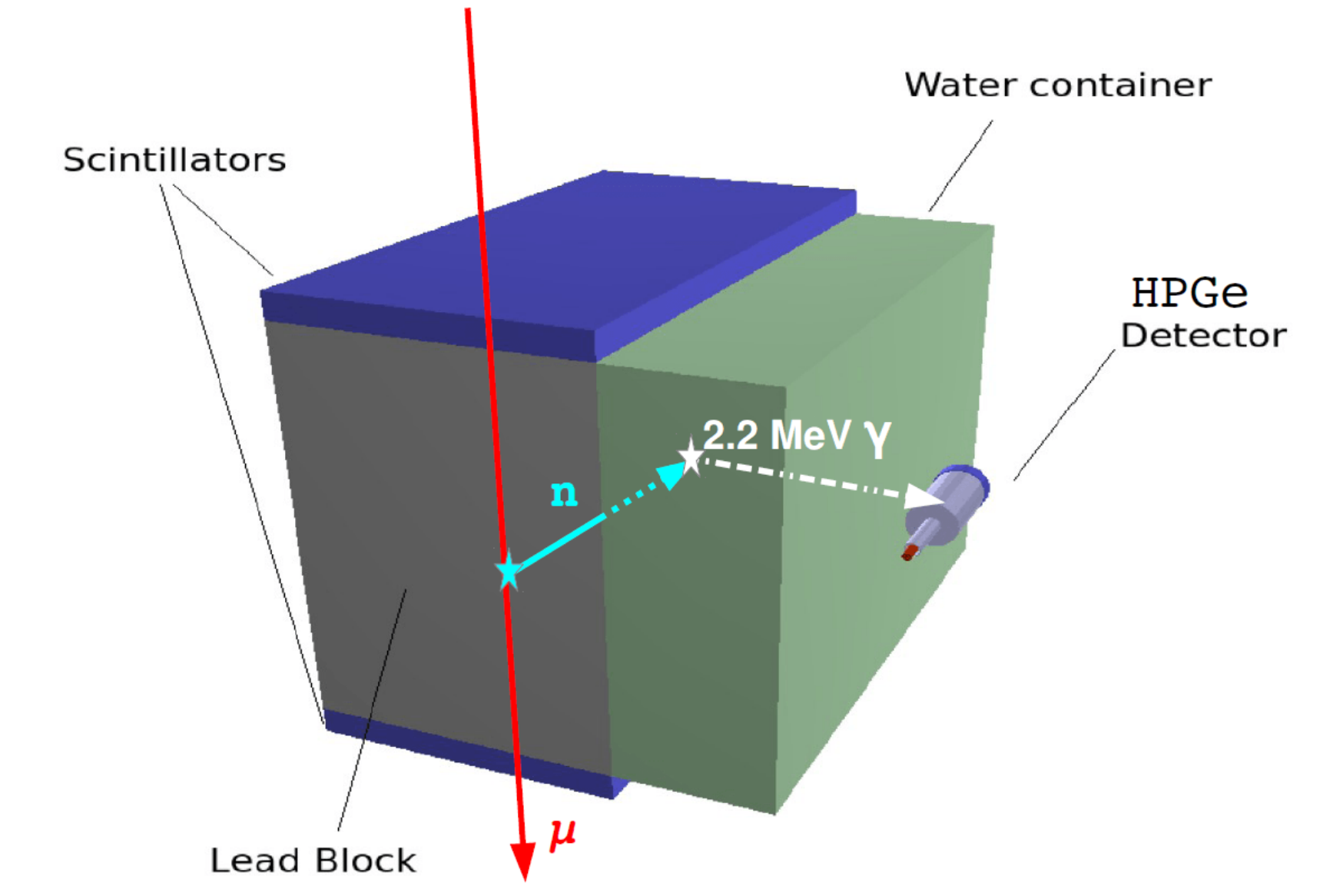}
  \end{center}
 \caption{Muon-induced neutrons are observed 
          indirectly through capture in water with 
          subsequent gamma ray emission. 
          Scintillators are used for muon tagging. 
          From \cite{palermo}.}
  \label{fig:minidex-working-principle} 
\end{figure}

The detection strategy, see also Fig.\,\ref{fig:minidex-working-principle},  
is:
\begin{itemize}
 \item muons passing through the high-Z material 
       are tagged with plastic scintillator panels;
 \item they interact with the high-Z target material;
 \item the muon-induced neutrons emerging from the high-Z volume 
       are thermalised 
       in water and eventually captured by hydrogen nuclei;
 \item the 2.2\,MeV gammas emitted after the neutron capture are identified 
       by germanium detectors;
 \item the signal is 2.2\,MeV gammas recorded within a predefined time 
       window after the passage of a muon;
 \item the background is measured between the end of the signal window 
       and the passage of the next muon. 
\end{itemize}
An important feature is that the background is measured by MINIDEX itself
without the need of any Monte Carlo calculation.

Indirect neutron detection through thermal capture in a 
selected isotope with subsequent gamma ray emission is a standard 
technique to detect neutrons~\cite{reichhart,kluck}.
The plastic scintillator panels tagging the muons
work independently from the germanium detectors identifying neutron
captures. 
Thus, there are no inefficiencies due to time resolution, which can be 
a problem for experiments using the same detectors to identify muons 
and neutrons. 

Lead was chosen as the target material for run\,I as it is one of 
the most commonly used materials for high-Z shielding  
in deep underground experiments~\cite{reichhart,majorana,exo200,cdex}.

\subsection{Run\,I geometry}
\label{sec:minidex-geometry}

\begin{figure}[!ht]
 \begin{minipage}[c]{.5\textwidth}
 \centering
    \includegraphics[width=1\textwidth]{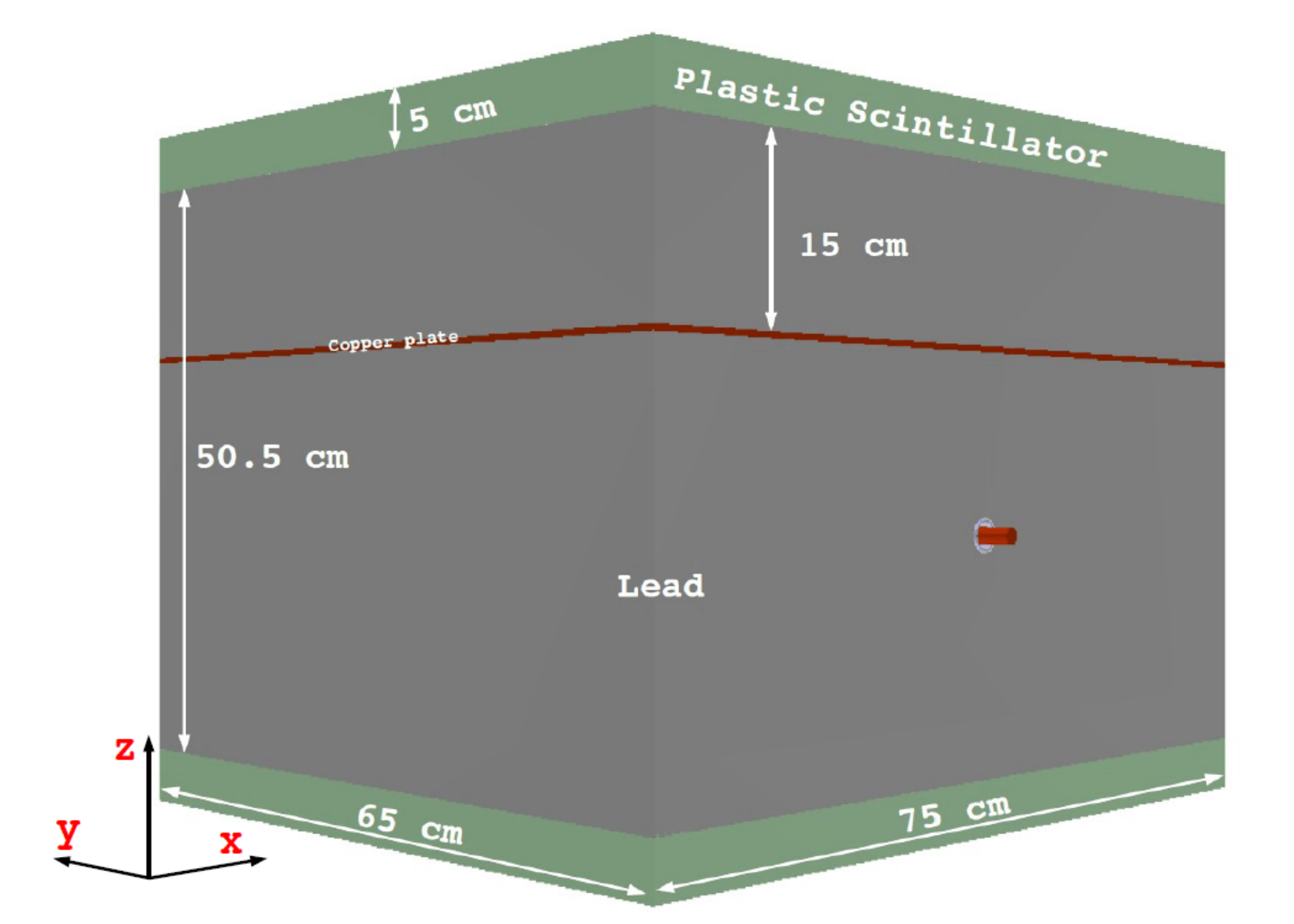}
 \subcaption{}
\end{minipage}
\begin{minipage}[c]{.5\textwidth}
 \centering
    \includegraphics[width=1\textwidth]{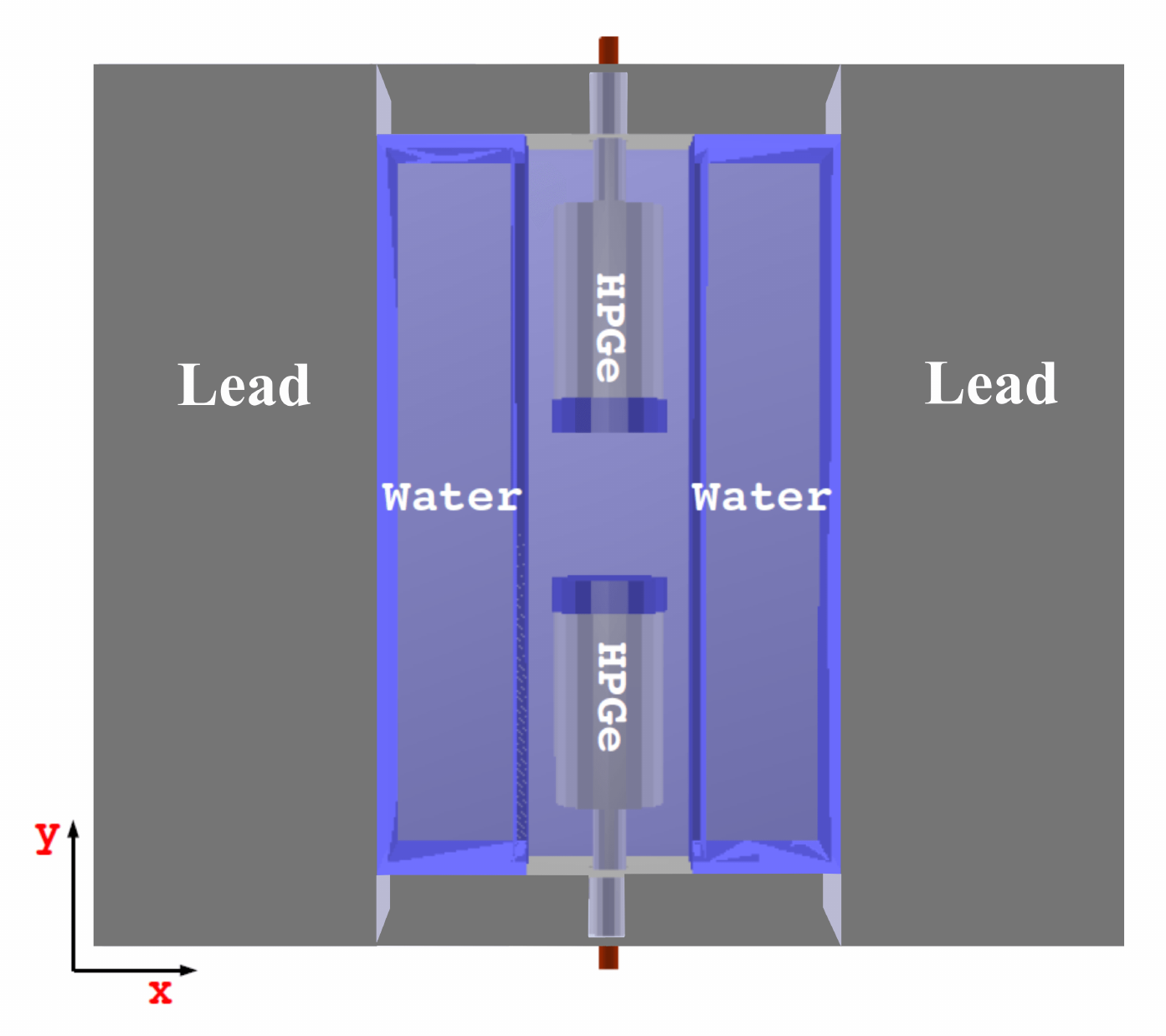}
 \subcaption{}
\end{minipage}
\caption{Schematic (a) of the outside of MINIDEX, 
          (b) central cut from the top of MINIDEX. 
          Both adapted from \cite{palermo}.
         }
\label{fig:sketch-total}
\end{figure}

MINIDEX
is a compact apparatus with a foot-print of 65\,$\times$\,75\,cm$^2$ and 
a height of 60.5\,cm. It consists of a lead castle with 
outer dimensions of 65\,$\times$\,75\,$\times$\,50.5\,cm$^3$. 
Plastic scintillator panels, 5\,cm thick,
exactly cover the lead surfaces
at the top and the bottom of the setup. 
The outer view of the MINIDEX apparatus is depicted 
in Fig.\,\ref{fig:sketch-total}(a). 

\begin{figure}[!ht]
 \begin{minipage}[c]{.5\textwidth}
 \centering
 \includegraphics[width=1\textwidth]{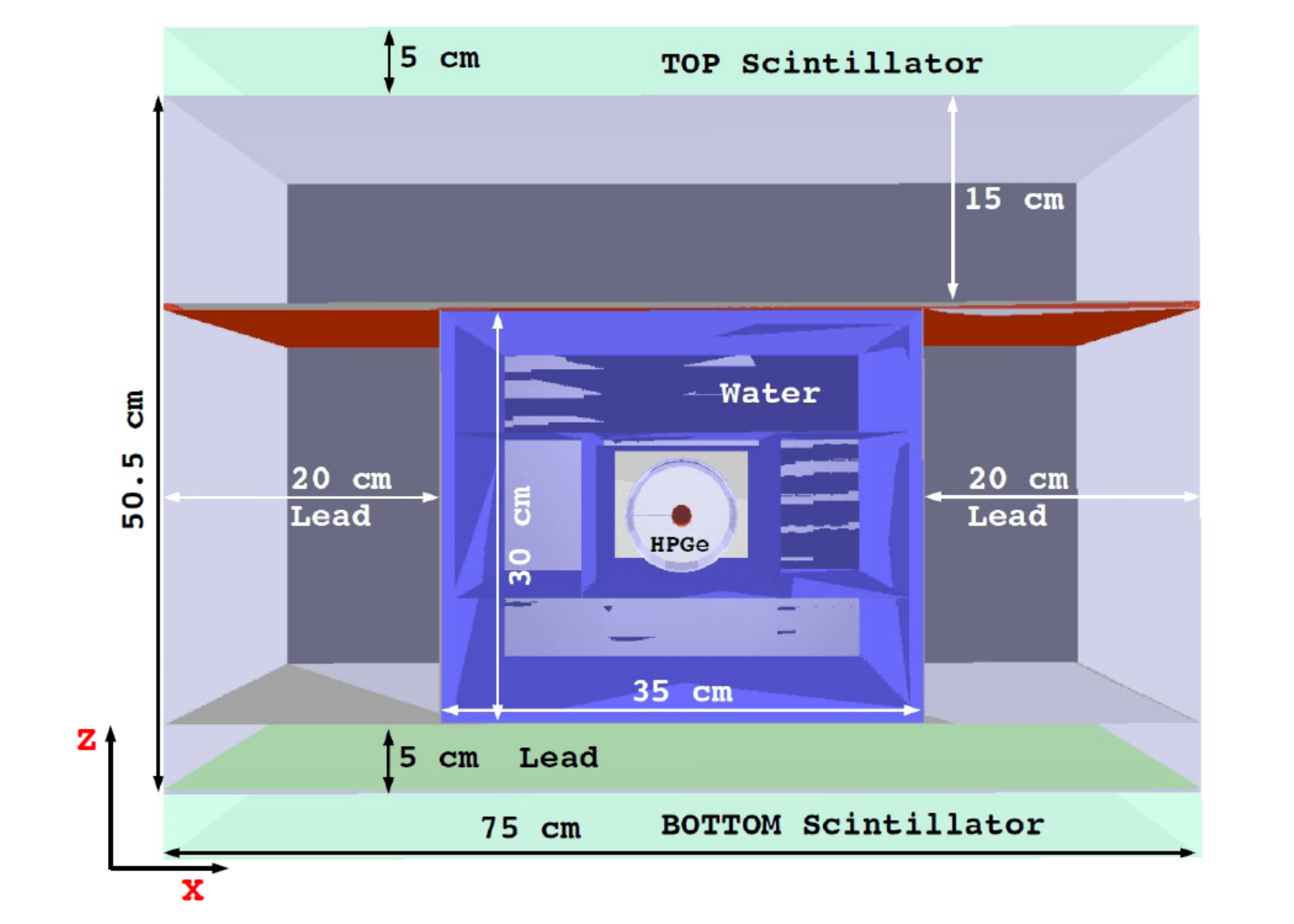}
 \subcaption{}
\end{minipage}
\begin{minipage}[c]{.5\textwidth}
 \centering
 \includegraphics[width=1\textwidth]{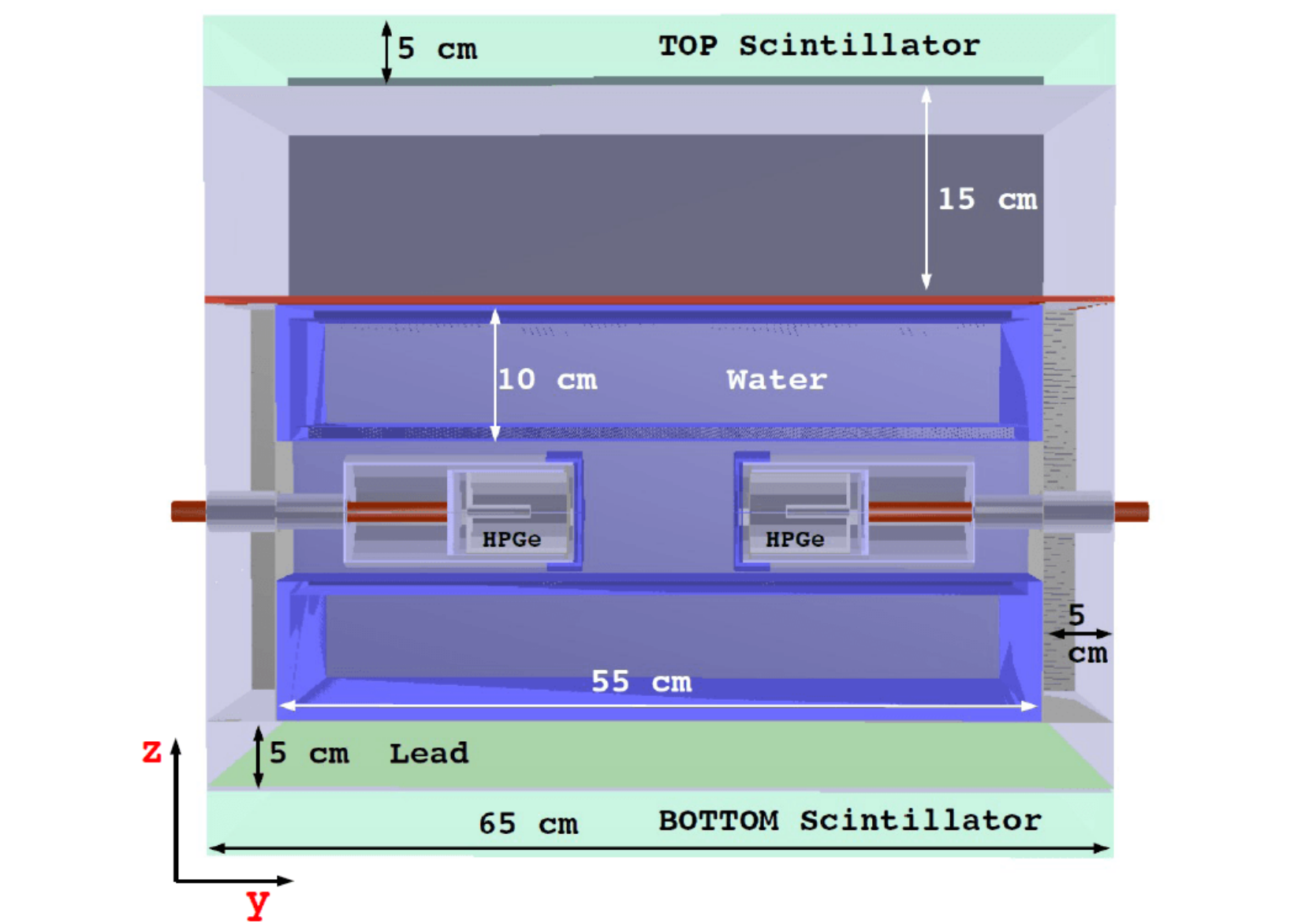}
 \subcaption{}
\end{minipage}
\caption{Schematic of central cuts through MINIDEX: (a) from the front,
 (b) from the side. Both adapted from \cite{palermo}.} 
\label{fig:sketch-sides}
\end{figure}

A rectangular container with outer dimensions 
35\,$\times$\,55$\times$\,30\,cm$^3$,
filled with ultra-pure water, 
is located inside the lead castle as shown in 
Figs.~\ref{fig:sketch-total}(b) and~\ref{fig:sketch-sides}.   
This container is made of plastic (C$_{10}$H$_8$O$_4$) and 
has a wall-thickness of 1\,cm.  
The minimum thickness of the water layer of 
about 9\,cm was optimized 
such that the probability to capture a neutron entering the water volume 
is about 95\,\%~\cite{palermo}.

The water container has a 
central hole which runs through the whole length of the container. 
It is 13\,cm wide and 8\,cm high. 
Two high-purity germanium (HPGe) detectors are placed in this hole, 
facing the center of the water container.  

\begin{figure}[!ht]
 \begin{minipage}[c]{.5\textwidth}
 \centering
 \includegraphics[width=1\textwidth]{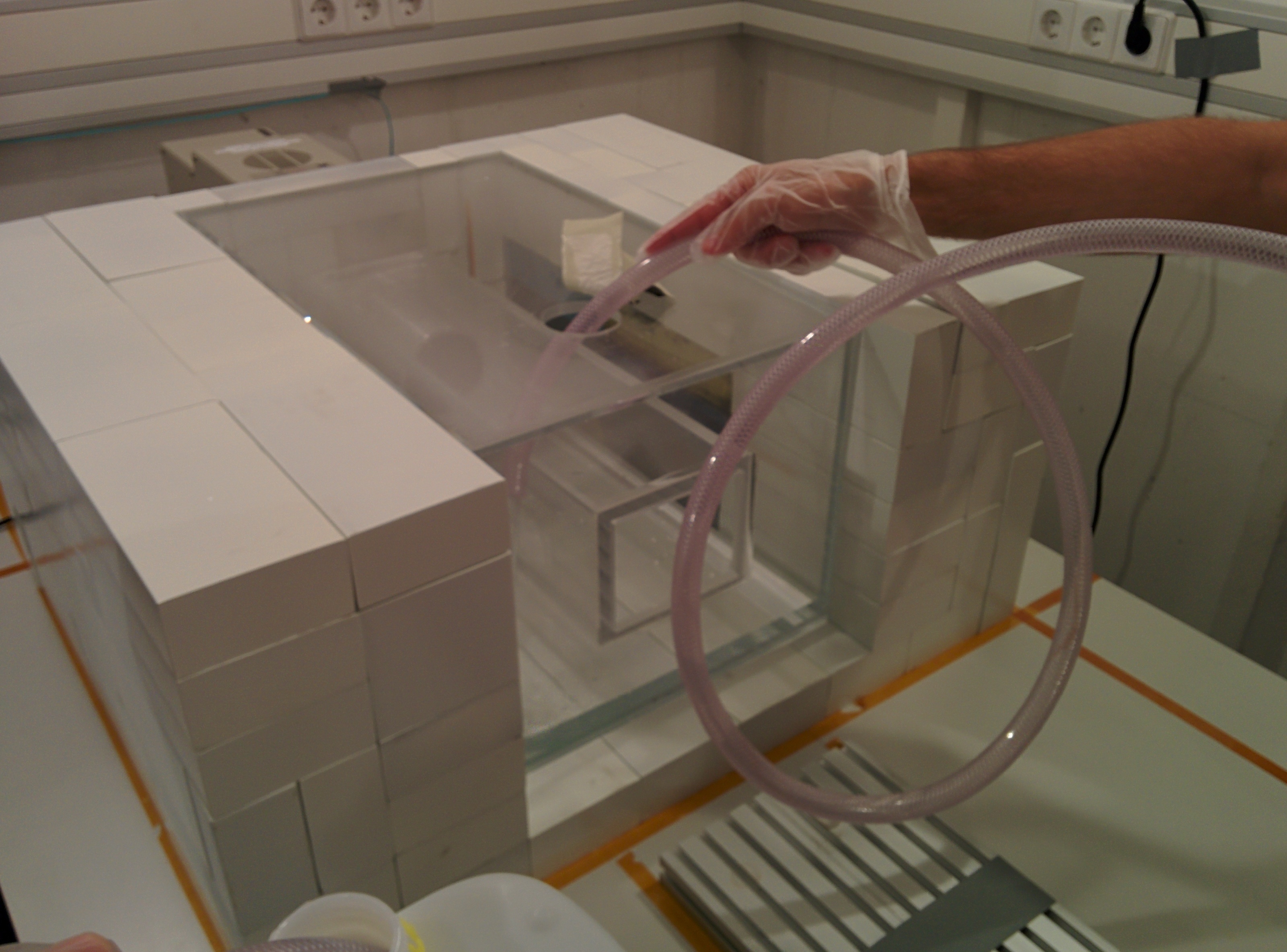}
 \subcaption{}
\end{minipage}
\begin{minipage}[c]{.5\textwidth}
 \centering
 \includegraphics[width=1\textwidth]{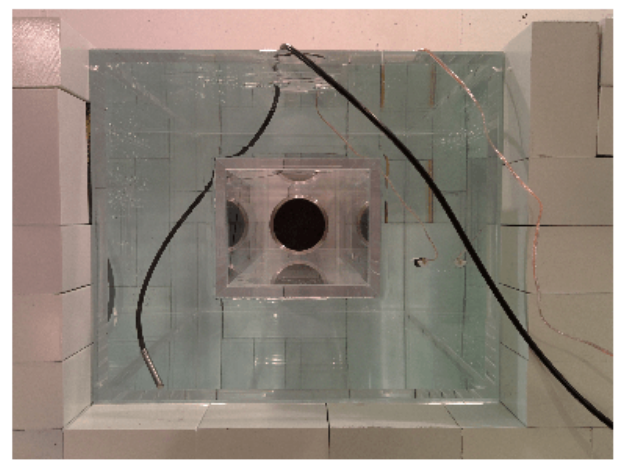}
 \subcaption{}
\end{minipage}
 \begin{minipage}[c]{.5\textwidth}
 \centering
 \includegraphics[width=1\textwidth]{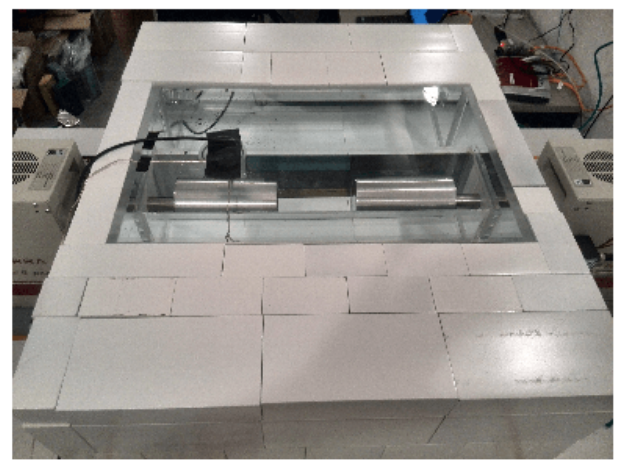}
 \subcaption{}
\end{minipage}
\begin{minipage}[c]{.5\textwidth}
 \centering
 \includegraphics[width=1\textwidth]{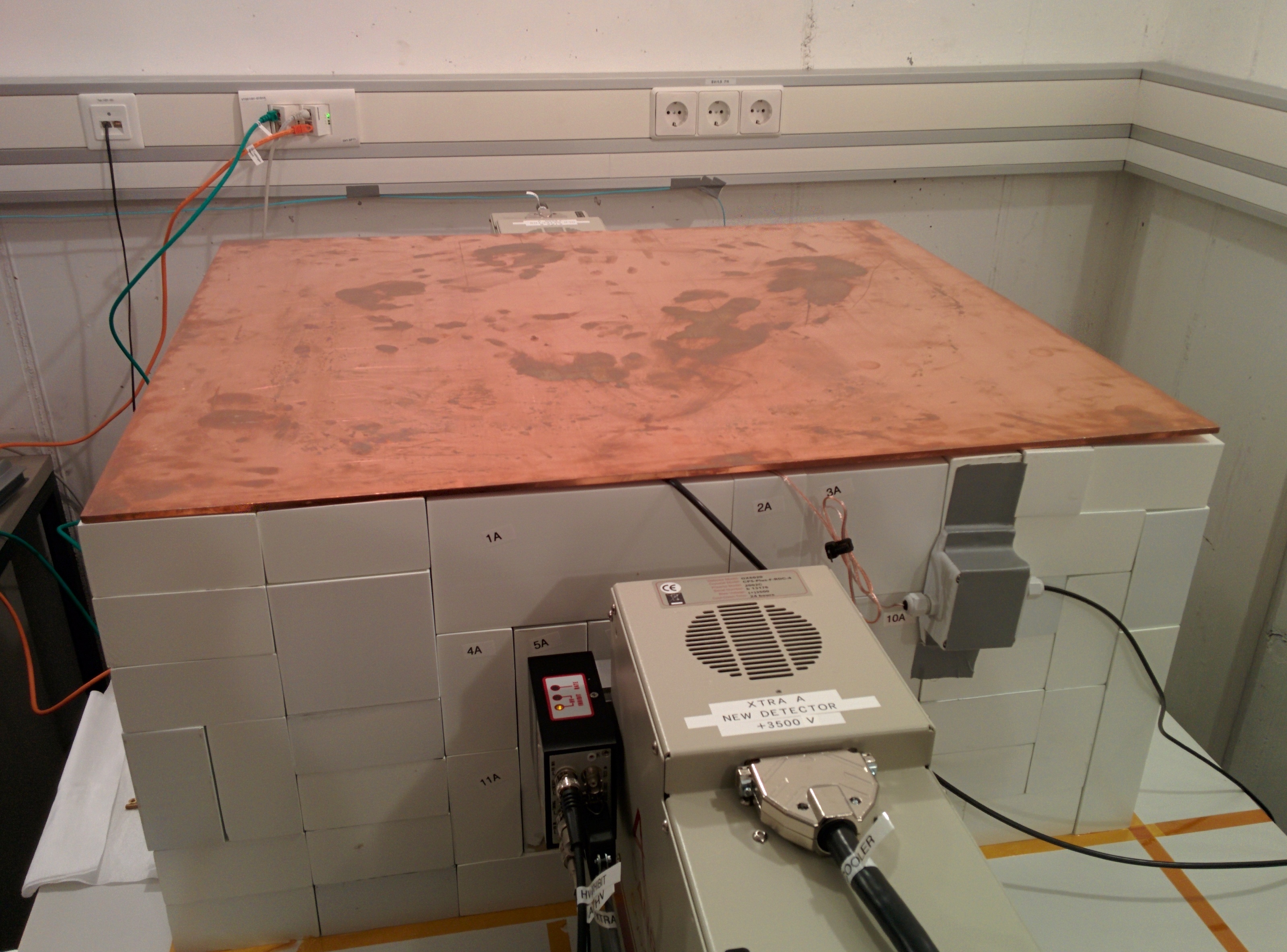}
 \subcaption{}
\end{minipage}
\caption{(a) Partially built up lead castle with water tank being filled;
         (b) frontal view into MINIDEX with one germanium detector inserted
             from the other side;
         (c) view from the open top onto the two germanium detectors;
         (d) the setup covered by the structural copper plate before
             stacking the lead on top.
          }
\label{fig:minidex-construction}
\end{figure}

As a support for the top layer of lead, there is a 0.5\,cm thick 
copper plate, see Fig.~\ref{fig:minidex-construction}(d). 
This copper plate avoids any weight load on the water container. 
Copper was chosen due to its relative low weight compared to its strength.

\begin{figure}[!h]
  \begin{center}
    \includegraphics[scale = 0.6]{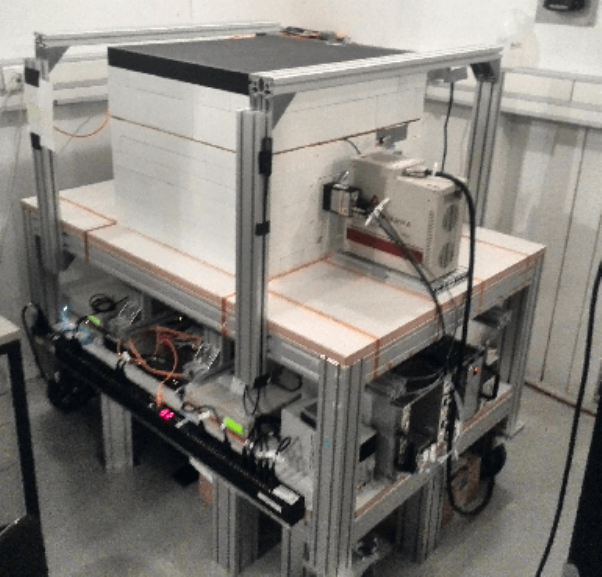}
  \end{center}
 \caption{The MINIDEX run\,I setup completely installed in the 
          T\"ubingen Shallow Underground Laboratory.
          Visible are the lead castle, the top scintillator panel,
          the electrical cooling unit of one of the germanium detectors
          and some electronic equipment on the lower shelf of the table.}
  \label{fig:minidex-full-view} 
\end{figure}

Photos of the MINIDEX setup during construction are shown 
in Fig.\,\ref{fig:minidex-construction}.
The lead castle is placed on top of a heavy-load table 
(surface area of 100\,$\times$\,150\,cm$^2$) with aluminum
framing as shown in Fig.~\ref{fig:minidex-full-view}. 
The bottom scintillator, supported by an aluminum structure,
is placed underneath the table top;
this is not indicated 
in Figs.~\ref{fig:sketch-total} and~\ref{fig:sketch-sides}. 
The lower shelf of the table is used for the data acquisition (DAQ) system 
and all electronic devices needed to operate MINIDEX.
The complete MINIDEX run\,I setup is shown in 
Fig.\,\ref{fig:minidex-full-view}. 
It was installed in 
the T\"ubingen Shallow Underground Laboratory in less than three days.

\newpage
\subsection{Physics composition of signal and background}

The signal for run\,I was defined as 2.2\,MeV gamma events
recorded in a predefined time window after the passage of a muon
through the top and bottom scintillators.
Muons going vertically through a side-wall had energies of
at least 1.2\,GeV; 
muons passing diagonally could have substantially lower energies.
Events were counted as signal independently of the location of the
muon interaction, which could not be determined in the run\,I geometry.
Some small number of events were counted as signal 
even though a muon only passed one scintillator panel,
because a secondary particle deposited energy in the bottom scintillator.

Background are events with a 2.2\,MeV gamma recorded due to
\begin{itemize} 
  \item neutrons induced by non-triggering muons; 
    \begin{itemize} 
      \item passing close by, but not entering the setup;
      \item passing only one of the scintillator panels
            due to geometrical reasons; 
      \item being stopped in the apparatus;
    \end{itemize} 
  \item cosmogenic neutrons;
    \begin{itemize}  
      \item due to the remnant cosmic neutron flux within TSUL;  
    \end{itemize} 
   \item neutrons from natural radioactivity; 
    \begin{itemize}  
      \item from fission;
      \item from ($\alpha$,n)-reactions. 
    \end{itemize} 
\end{itemize}

\subsection{Detectors}

The MINIDEX run\,I setup requires only four detectors to be read out: 
two plastic scintillator panels and two HPGe detectors.


The two scintillator panels for run\,I, 
produced by Saint-Gobain Crystals \cite{saint-gobain}, 
were made of BC-408 (Polyvinyltoluene, C$_{10}$H$_{11}$)
with a density of 1.032\,g/cm$^3$. 
The wavelength of maximum emission was $\approx$ 425 nm.
Both the photomultiplier tubes (PMT) and the PMT HV bases (model HV2520AN) 
were embedded inside the panel volumes and no wavelength shifters were 
used to collect the light of the scintillators onto the PMT photocathodes. 
The PMTs had a diameter of 30\,mm and were
produced by ET Enterprises~\cite{ET-enterprises}(model 9900B). 
The active diameter was 25\,mm and the spectral range 
was 280\,nm to 680\,nm, with a peak quantum efficiency of 26\,\% 
at around 400\,nm. 
Sensitive PMT sidewalls allowed for wide-angle light detection. 
The tube housing the multiplication dynodes was $\approx 9$\,cm long.
  
The novel approach of having the PMTs incorporated in the
scintillator panels was chosen as it is particularly space saving.
However, the panels proved to be not fully efficient over the whole
surface area. The top and bottom panels were efficient 
to 86.53\,$\pm$\,0.10 and 92.69\,$\pm$\,0.15\,\%, 
respectively\,\footnote{These 
efficiencies  were unexpectedly low. They cannot be explained
by the geometrical effects due to the embedding  of the PMTS.
The PMT sidewalls were probably less sensitive than expected. 
A detailed investigation revealed that the PMTs were not properly 
optically coupled. The panels were replaced by Saint-Gobain and
exchanged for conventional panels with external PMTs after run\,I.}.
The data were corrected for these inefficiencies in the analysis.


The two HPGe detectors used are commercial Extended Range (XtRa) 
coaxial germanium detectors, produced by CANBERRA~\cite{canberra}. 
Their nominal energy resolution is $\approx 2$\,keV at 1.3\,MeV. 
These detectors are manufactured 
from  cylindrical germanium crystals with radii of 3.5\,cm,  
lengths of $\approx 7$\,cm and central bore holes with a length of 4.5\,cm. 
The crystals are housed in cylindrical vacuum chambers made of aluminum. 
The detectors are in thermal contact with copper cooling fingers 
which extend beyond the lead castle to the electrical cooling 
units, see Figs.~\ref{fig:sketch-sides}(b)
and~\ref{fig:minidex-full-view}.  
Having electro-cooled germanium detectors makes MINIDEX an
almost maintenance-free setup.

\newpage
\subsection{DAQ and Electronics}

The DAQ system employed for MINIDEX is a 
16-channel VME digitizer card, SIS3316-DT, produced by 
Struck Innovative Systems~\cite{struck}. 
Each channel can be used independently from the 
others with a 250 MHz sampling rate. 
All channels are equipped with a 14-bit resolution ADC 
and run in double bank mode to avoid dead time. 
The time information is distributed to the channels via an internal clock. 

All detector signals from MINIDEX are recorded independently 
without a shared trigger. 
Data are written to and stored in a server 
(sysGen/Supermicro SYS-5018D-MTF~\cite{supermicro}) 
which is placed directly next to the DAQ on the lower shelf of the 
supporting table.

A two-channel HV power supply (iseg NHQ 206L) provides the high voltage 
for the HPGe detectors. Low voltage supplies support 
the preamplifiers of the germanium detectors and the scintillator panels.
The whole system is being controlled remotely through a router, 
which provides access to every single device.
To prevent damage to the germanium detectors from power cuts 
and to generally protect the system from noise due to the power line,
MINIDEX is operated behind  an uninterruptible power supply 
(Online XSR3000 PSU).

\subsection{Online Monitoring and Detector Performance}

The MINIDEX apparatus is continuously monitored during operation. 
The energy spectra of the scintillators and the HPGe detectors are
recorded and the data quality is constantly evaluated through the 
energy resolution of both HPGe detectors at 1460\,keV ($^{40}$K line). 
Also recorded are the temperature of the water and its level inside the tank. 

\begin{figure}[!h]
  \begin{center}
    \includegraphics[width=1\textwidth]{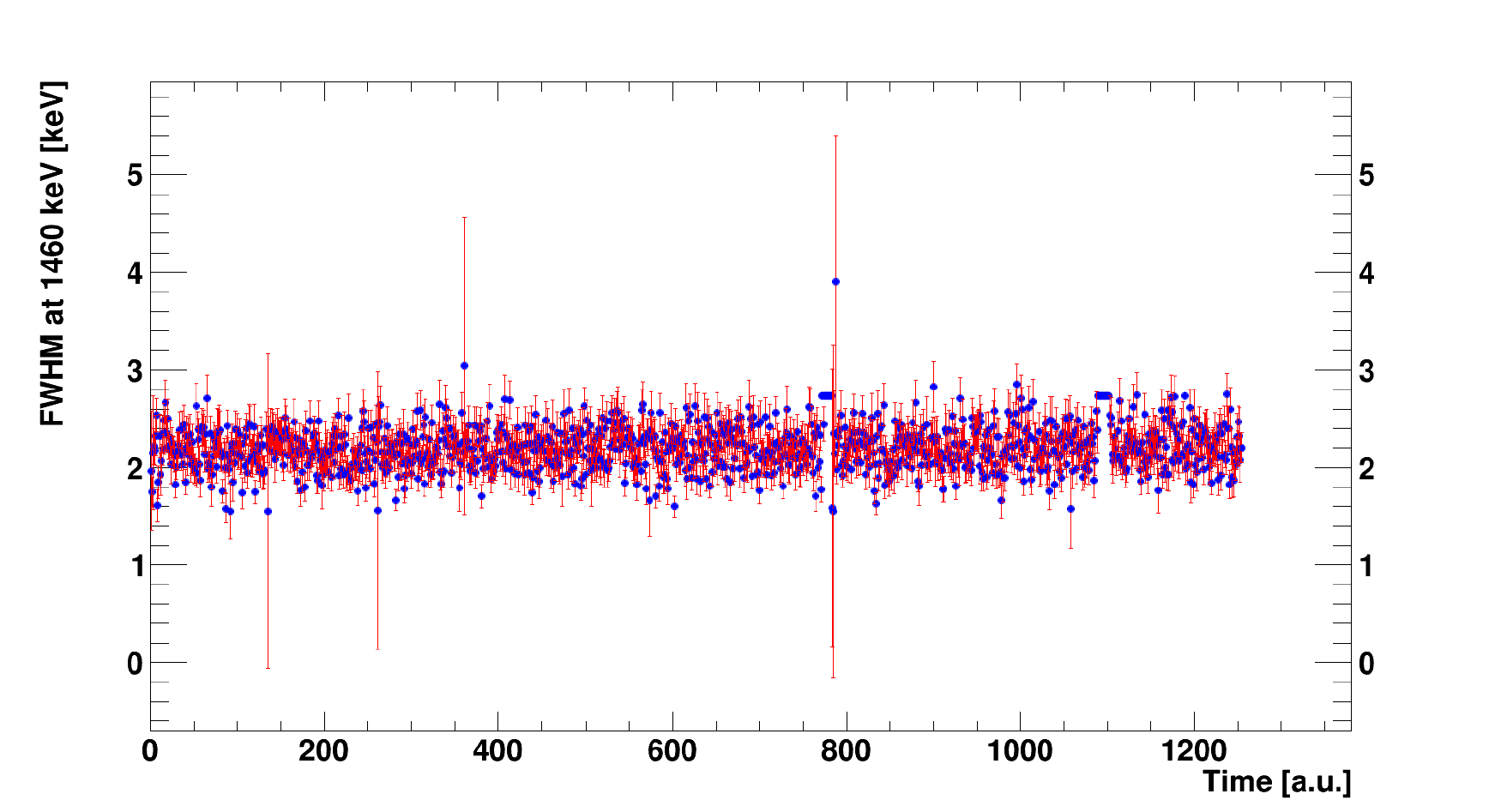}
  \end{center}
 \caption{The energy resolution at 1460\,keV of the sum of the spectra of 
          the two germanium detectors vs. time. Each point corresponds to 
          approximately two hours of data taking. All data collected 
          between the beginning of MINIDEX run\,I and November, 25, 
          2015 are included. From \cite{palermo}.}
  \label{fig:en-resolution-stability} 
\end{figure}

Figure\,\ref{fig:en-resolution-stability} demonstrates the stability 
of the energy resolution of 2.2\,keV at 1460\,keV for 
the sum of the energy spectra of the two germanium detectors. 
Each point represents approximately two hours of data taking. 

\begin{figure}[!ht]
 \begin{minipage}[c]{.5\textwidth}
 \centering
 \includegraphics[width=1\textwidth]{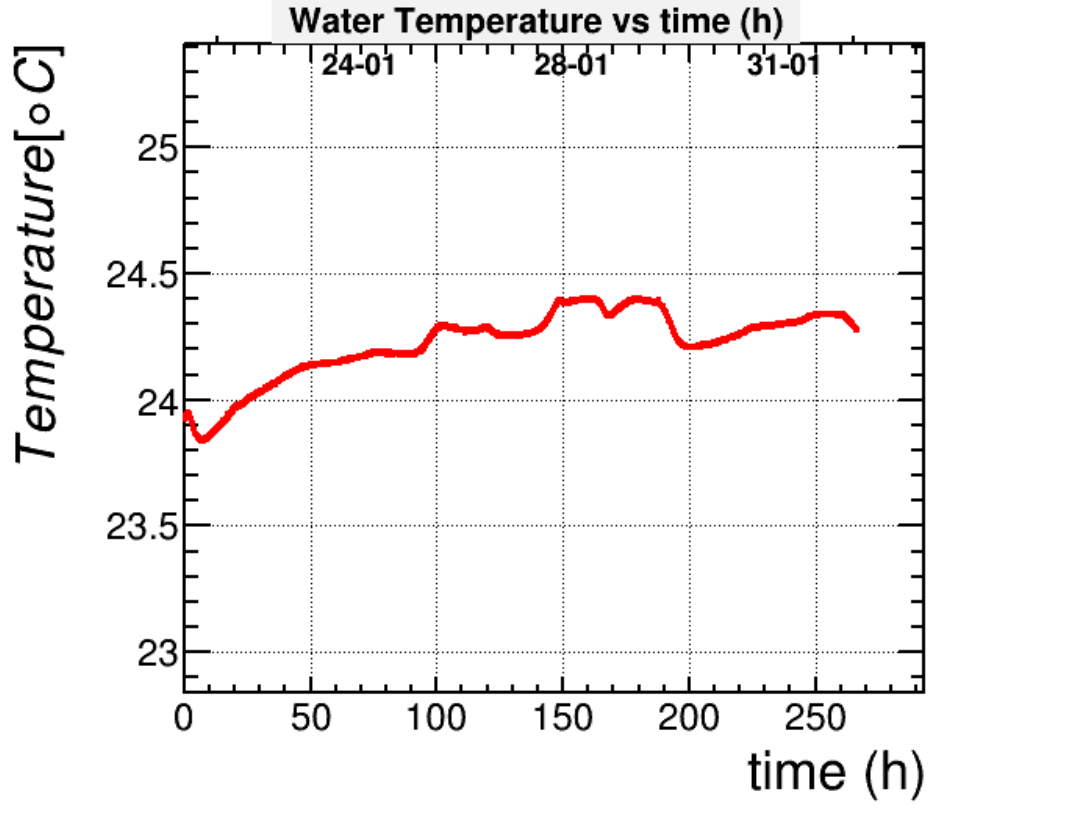}
 \subcaption{}
\end{minipage}
\begin{minipage}[c]{.5\textwidth}
 \centering
 \includegraphics[width=1\textwidth]{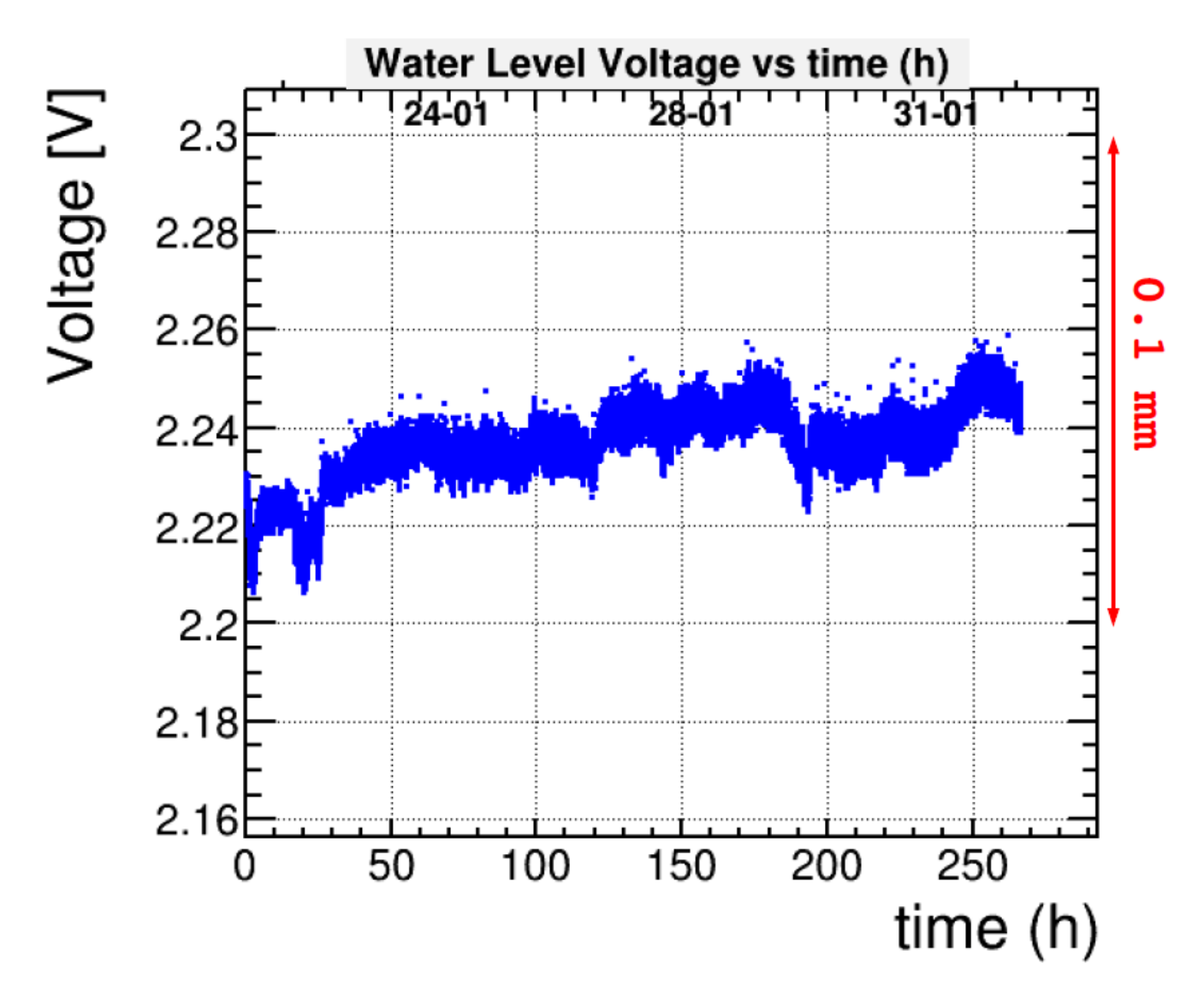}
 \subcaption{}
\end{minipage}
\caption{(a) Water temperature vs. time. 
         (b) Water level vs. time;
             the signal is a voltage -- a difference of 0.1 V  
             corresponds to a change in water-level of 100\,$\mu$m. 
             Both adapted from \cite{palermo}.} 
\label{fig:minidex-temp-level}
\end{figure}

The temperature of the water was measured with a PT100~\cite{pt100} 
temperature sensor placed inside the water tank. To measure the water level 
inside the tank, MINIDEX was equipped with a liquid-level sensor probe 
produced by Vegetronix~\cite{vegetronix}. The stability of both temperature 
and water level is demonstrated in Fig.~\ref{fig:minidex-temp-level}. 
They vary only very little and the change in level is compatible
with the expansion of the water due to temperature shifts.
Thus, the mass of the water was constant; there were no leaks
over the whole data taking period.

\section{Run\,I}

Run\,I started on July 15, 2015, and the setup was
operated in a stable configuration until January 19, 2016.
The data taken until November 25, 2015 were 
included in the first analysis. 
They represent a run-time of $9.8 \times 10^6$\,s (114.4 days).
This relates to an up-time of 86\,\%~\footnote{The down-time was caused by the
infant mortality of HV supplies for the scintillators.}.

\subsection{Offline trigger definition}
\label{sec:analysis-strategy}

\begin{figure}[!h]
  \begin{center}
    \includegraphics[scale = 0.35]{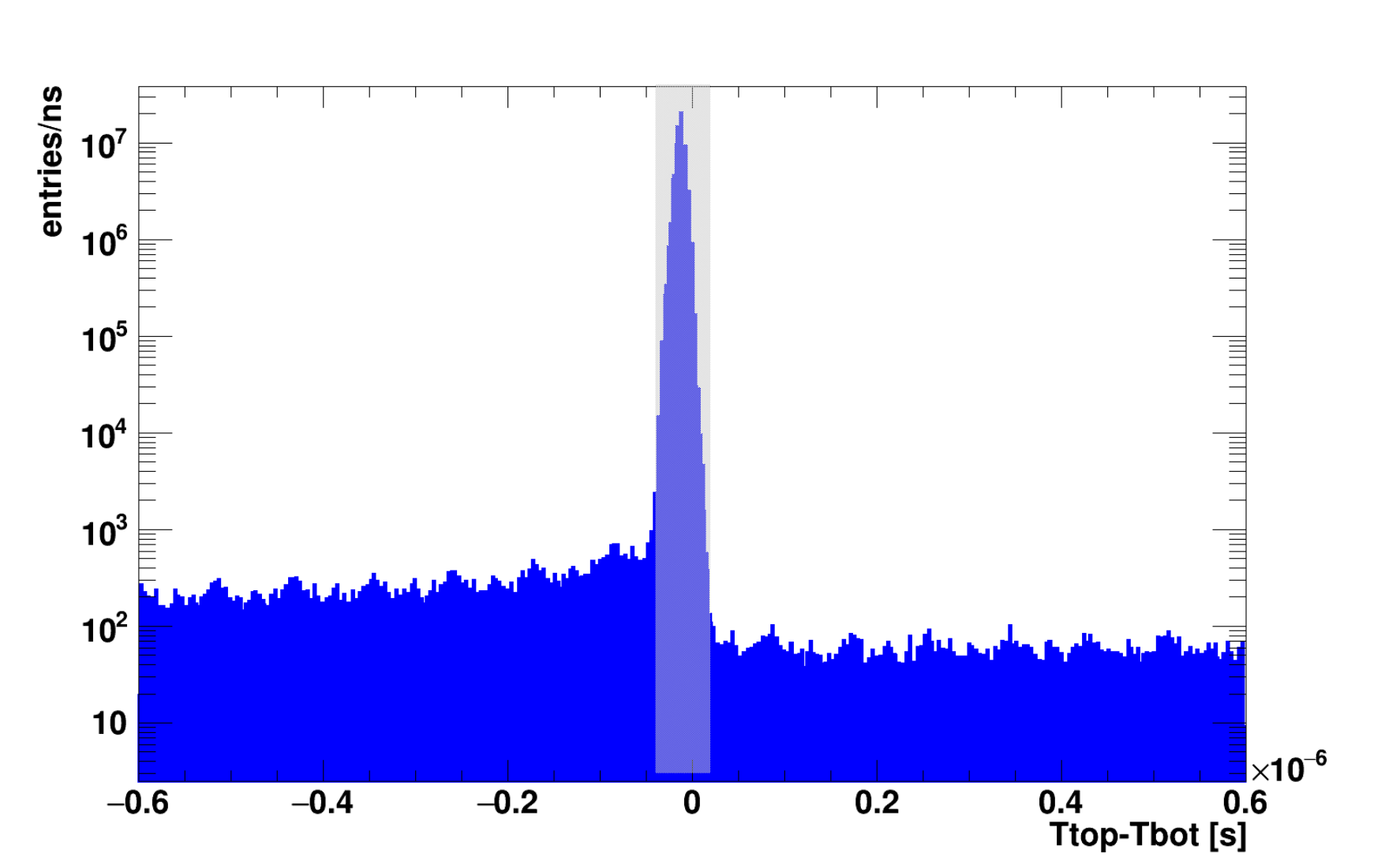}
  \end{center}
 \caption{Distribution of the time difference $T_{\rm top}-T_{\rm bot}$.
          The value of a given $T_{\rm top}$ is compared to the previous
          and all following $T_{\rm bot}$ until a new $T_{\rm top}$ is recorded.
          The lightly shaded area around the pronounced peak 
          corresponds to the trigger window. From \cite{palermo}.}
  \label{fig:tdistr} 
\end{figure}

The trigger for run\,I requires a muon passing through
both the top and the bottom scintillator panels.
Such muons are identified through signals in both panels within
a time window to be determined.
The signals from the top and bottom scintillators are recorded 
independently at times denoted
$T_{\rm top}$ and $T_{\rm bot}$, respectively. 
The distribution of the time difference $T_{\rm top}-T_{\rm bot}$ is 
shown in Fig.~\ref{fig:tdistr}.
A very significant peak characterises the coincidences
associated with the passage of muons.  
The width of the peak is related to time jitter.
The trigger window chosen was  $[-40;+20]\,\mbox{ns}$.
For each triggered event, $T_{\rm top}$ was taken as the time 
of the passage of the muon, $T_{\rm trig}$.
No cut on the light observed in each panel was 
applied\,\footnote{As the spectra of the panels did not show a 
good separation between single hits and coincidences, this was
unavoidable.}.

Approximately $10^8$\,triggers were identified,
corresponding to a trigger rate of $\approx 10$\,Hz.  
The individual total count rates of the scintillator panels 
were ${R_{\rm top}} = 49.3$\,Hz and  ${R_{\rm bot}} = 46.8$\,Hz, respectively. 
Given the width of the coincidence window of $\Delta T_{coinc}$ = 60 ns, 
the corresponding accidental trigger rate of $\approx 3 \times 10^{-4}$\,Hz
only produces a negligible contamination of the triggered events.
Outside the trigger window, an excess of events with negative  
$T_{\rm top}-T_{\rm bot}$ was observed. This is due to muons only passing through
the top scintillator and secondary particles depositing energy 
in the bottom scintillator. For muons only passing through the bottom scintillator,
it is less likely that secondary particles reach the top scintillator.

\subsection{Definition of signal window}

The data recorded for the two HPGe detectors are divided into two time windows
after each trigger:
\begin{itemize}
 \item \emph{INSIDE}: events recorded  within a time window  $\Delta T_{win}$;  
 \item  \emph{OUTSIDE}: events recorded between the end of $\Delta T_{win}$ 
   and the arrival of the next trigger.
\end{itemize}

The value of  $\Delta T_{win}$ is not fixed, i.e.\ the analysis is
performed for multiple  $\Delta T_{win}$ to study the time evolution 
of the signal.
Simulations performed for the design studies 
predicted that the time distribution 
of the muon-induced 2.2\,MeV gammas seen by the germanium detectors would not 
extend significantly beyond  1\,ms after the occurrence of a trigger.
Thus, the average time between triggers of $\approx 100$\,ms is 
large enough to not affect the full collection of the signal.

\begin{figure}[!h]
  \begin{center}
    \includegraphics[scale = 0.5]{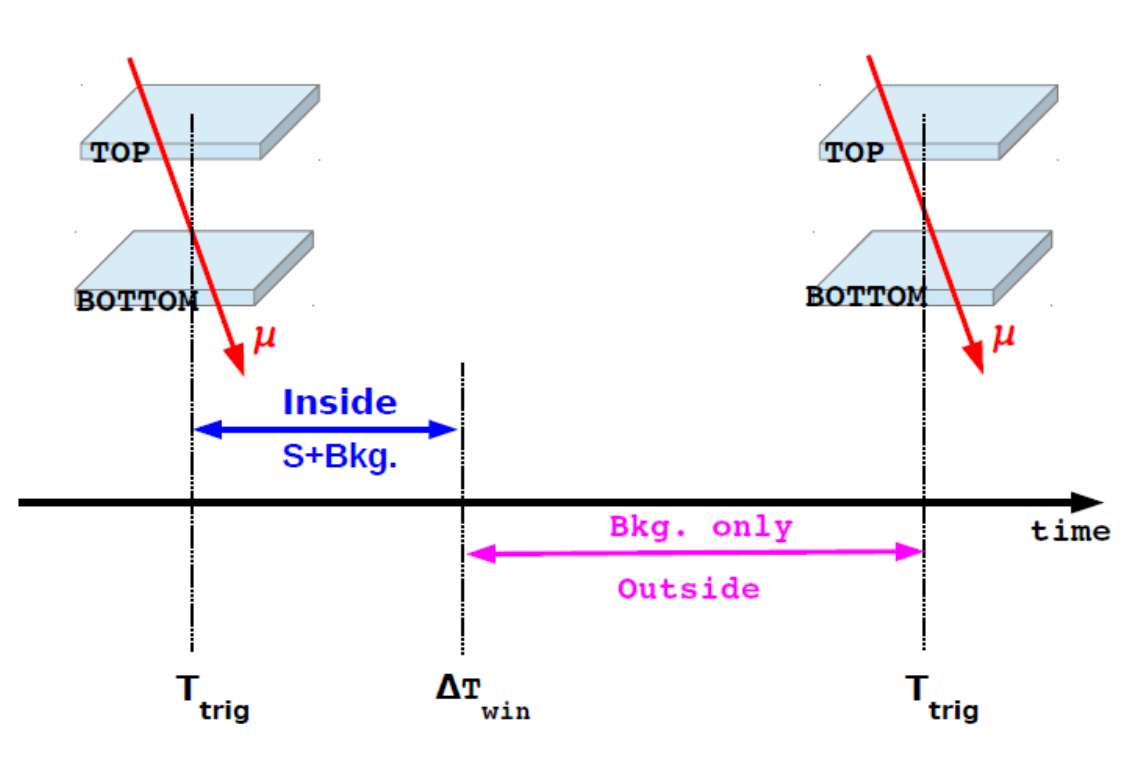}
  \end{center}
 \caption{Sketch illustrating the analysis strategy adopted for
          MINIDEX run\,I. When $\Delta T_{win}$ is big enough, the outside
          window  contains only background (Bkg.) events. 
          The inside window contains also the full signal(S). 
          From \cite{palermo}.}
  \label{fig:analysis-strategy} 
\end{figure}

With large enough $\Delta T_{win}$,
the inside window fully contains the signal plus background while
the corresponding outside window only contains events 
related to background.
Thus, the background is measured in parallel to the signal.
This procedure is  illustrated in Fig.~\ref{fig:analysis-strategy}.

\subsection{The germanium detector signal}

\begin{figure}[!ht]
  \begin{center}
    \includegraphics[scale = 0.45]{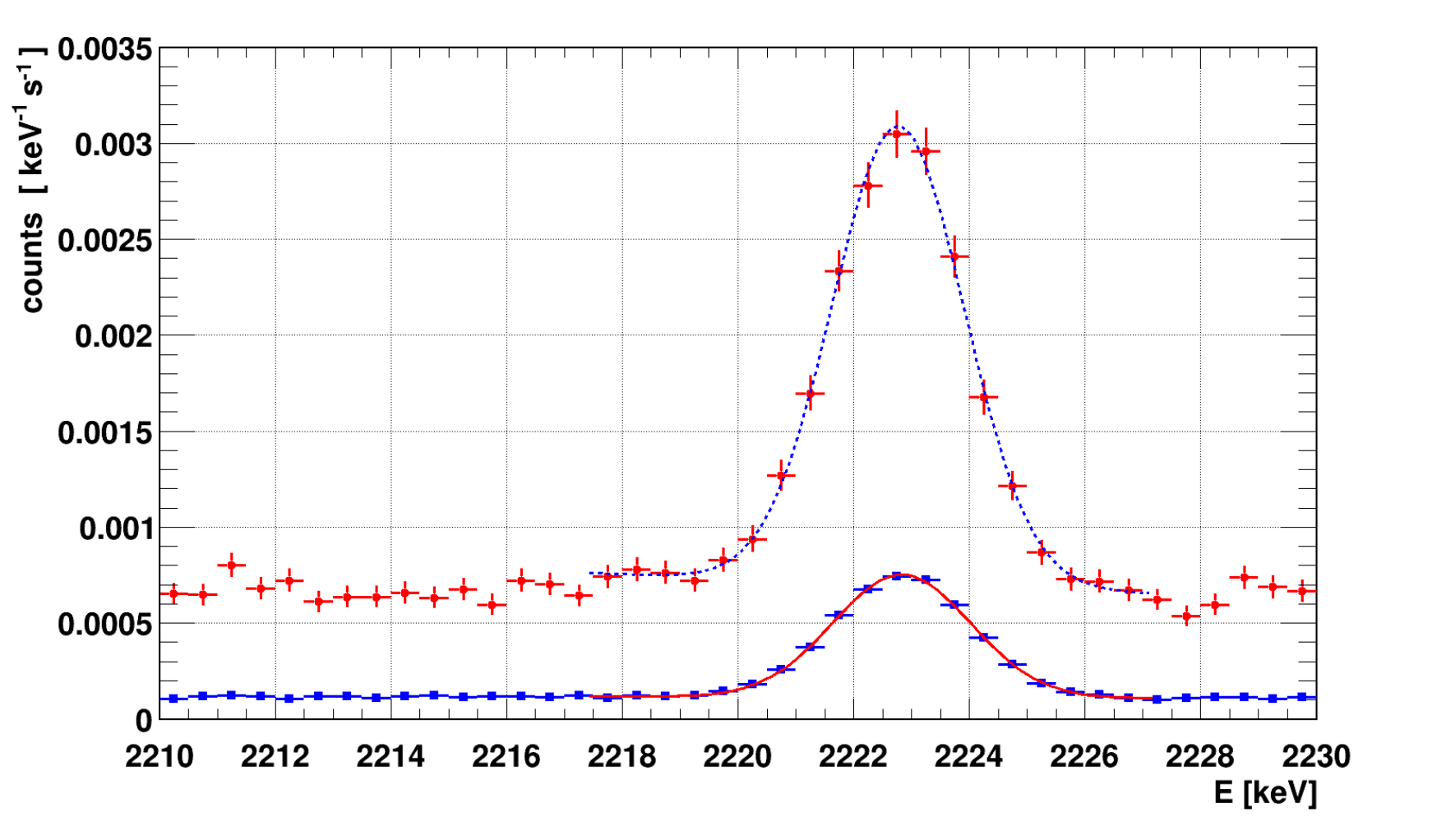}
  \end{center}
\caption{Summed time-normalised spectra of the two germanium detectors 
         for inside (circles) and outside (squares) the 
         $\Delta T_{win} = 4$\,ms window for the complete run-time.
         Both spectra were fit with a Gaussian plus a first order
         polynomial. The fit results are shown as a dotted (solid) line
         for the data inside (outside) the time window.
         The error bars represent the statistical uncertaintenties.
         For the outside spectrum, the statistical uncertainties are smaller
         than the size of the symbols.
         }
\label{fig:ger-signal}
\end{figure}

The calibrated spectra of the germanium detectors were summed up
for the inside and outside windows.
Figure~\ref{fig:ger-signal} shows the result for $\Delta T_{win} = 4$\,ms.
Also shown in Fig.~\ref{fig:ger-signal} are fits with a Gaussian plus
a first order polynomial.
The resolution of the 2.2\,MeV peak is 2.8\,keV FWHM.
The contribution of neutron induced 2.2\,MeV gammas
is measured over a much larger time for the background than for the signal and
thus, to a much higher precision. This is important for background 
subtraction.

The numbers of neutron-induced events,
${N_{peak}}^{inside}$ and ${N_{peak}}^{outside}$, 
were evaluated from fits as shown in Fig~\ref{fig:ger-signal}.
To evaluate the systematic uncertainty on the result of the extraction, 
a method based on side-bands around a region of interest 
of 2.223\,MeV\,$\pm$\,5\,keV was used\,\cite{palermo}.
Any difference between the methods was taken as a symmetric systematic
uncertainty and added in quadrature to the statistical uncertainty.
The systematic uncertainties are in general much smaller than
the statistical uncertainties. They are only significant for very small
$\Delta T_{win}$. Unless otherwise stated, total uncertainties are quoted.

\subsection{Evaluation of signal and background rates}

The signal is determined through the rate of 2.2\,MeV gammas recorded
in the predefined time window after a  muon trigger. 
The effective run-times accumulated in the signal
window, ${ET_{inside}}$,
and the background window , ${ET_{outside}}$, are defined as
\begin{eqnarray}
{ET_{inside}} & = & {N_{trig}} \cdot \Delta T_{win} \mbox{ ,} \\
{ET_{outside}} & = & RT - {ET_{inside}} \mbox{ ,}
\label{eqn:ET-inside}
\end{eqnarray} 
where ${N_{trig}}$ is the number of triggers\,\footnote{Due to the inefficiency
of the scintillators, some muons were not triggered and thus the resulting
neutrons were not recorded as signal. This was corrected for by 
adjusting the number of triggers appropriately~\cite{palermo}.}
and RT the total run-time of the measurement. 

The effective rates for signal plus background, $\Gamma_{S+B}$, 
and background, $\Gamma_{B}$, are defined as
\begin{eqnarray}
{\Gamma_{S+B}} & = & {N_{peak}}^{inside} / {ET_{inside}} \mbox{ ,} \\
{\Gamma_{B}} & = & {N_{peak}}^{outside} /  {ET_{outside}}  \mbox{ .}
\label{eqn:rate-inside}
\end{eqnarray}

\begin{figure}[!h]
  \begin{center}
    \includegraphics[width=0.9\textwidth]
               {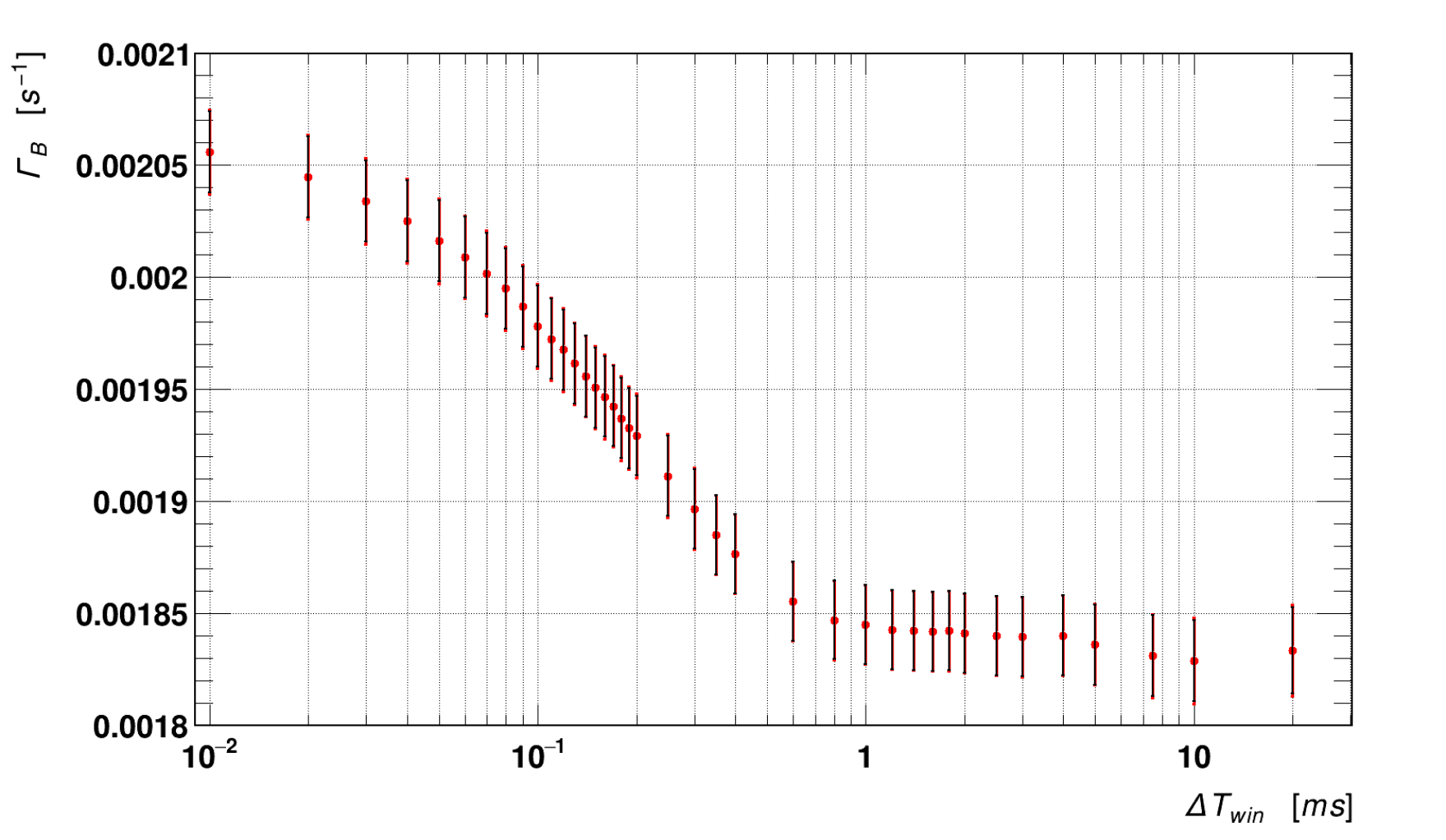}
  \end{center}
 \caption{The effective rate, ${\Gamma_B}$, of events observed in 
          the outside window.
  The inner error bars represent the statistical uncertainties, the
  outer bars represent the statistical and the systematic uncertainties 
  added in quadrature. The statistical uncertainties dominate such that
  the outer error bars are almost invisible. The entries are highly
  correlated as the events for each window also contain the events in the
  smaller time windows.}
  \label{fig:B-rate-data-fit-with-systematic-error} 
\end{figure}

The background rate,  ${\Gamma_{B}}$, was evaluated
for 39 values of $\Delta T_{win}$, see
Fig.~\ref{fig:B-rate-data-fit-with-systematic-error}.
The increase per step in $\Delta T_{win}$ is 10\,$\mu$s up to 200\,$\mu$s
and then the step size gradually gets larger.
The background rate is expected to be time independent.
For values of $\Delta T_{win} < 1$\,ms, a significant number of
signal events are ``leaking'' into the background window. 
As expected, the rate flattens out at $\Delta T_{win}$ values of around 1\,ms. 
To determine the overall background rate, $R_B$, safely, 
$\Delta T_{win}$ should be larger than 1\,ms.
A value of $\Delta T_{win} = 4$\,ms was chosen and $R_B$ determined to be
$R_B = (1.84 \pm 0.02) \times 10^{-3}$\,Hz.
The effective signal rate, $R_S$, for all predefined signal windows 
was calculated as 
\begin{equation}
R_S = {\Gamma_{S+B}} - R_B ~~.
\label{eq:Rs}
\end{equation}

\subsection{Time evolution of the signal}

\begin{figure}[h]
  \begin{center}
    \includegraphics[width=0.8\textwidth]
   {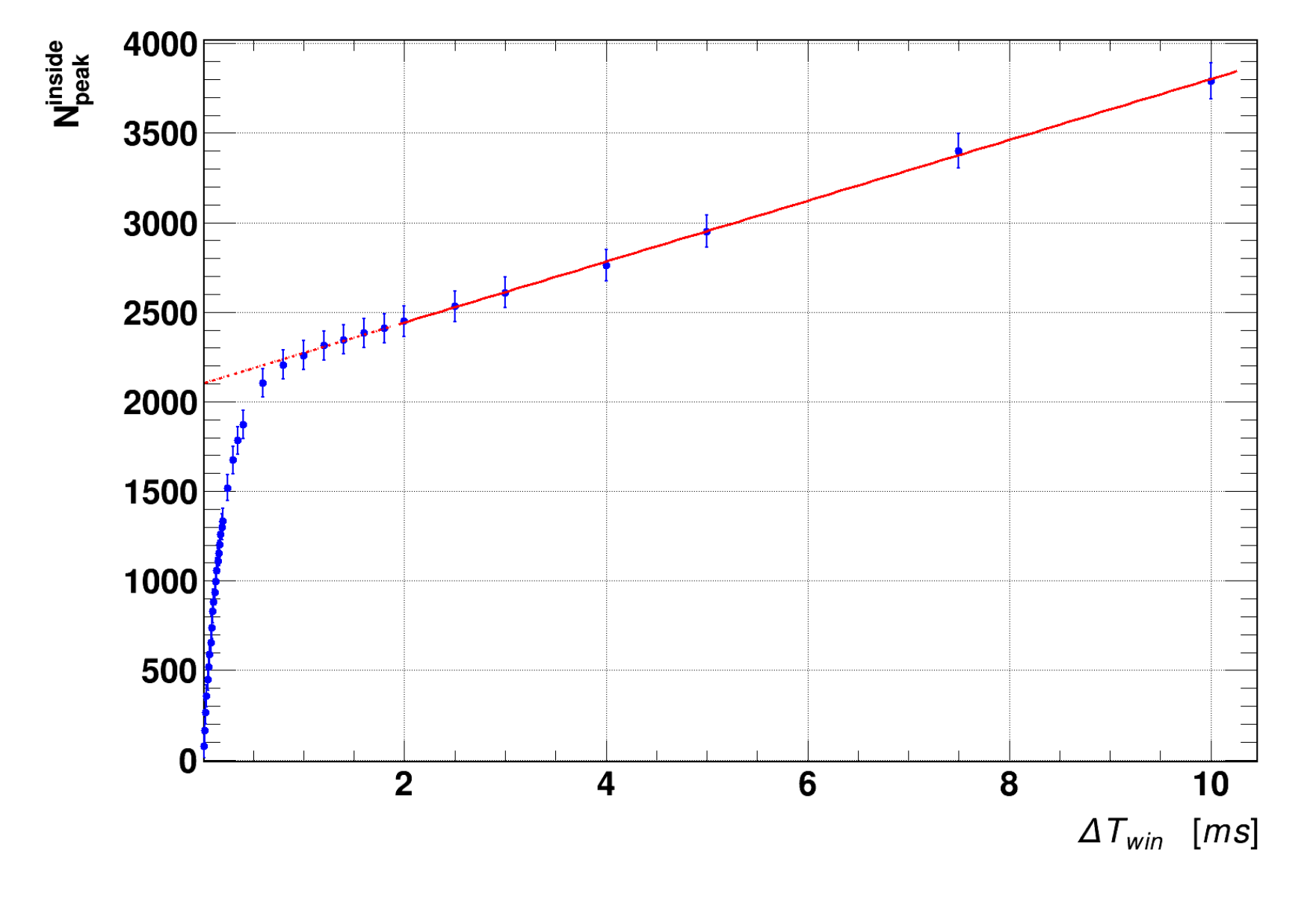}
  \end{center}
 \caption{Number of events,${N_{peak}}^{inside}$, 
          accumulated vs.\ $\Delta T_{win}$.
          A linear fit, represented by a solid line,
          was performed for $\Delta T_{win} \ge 2$\,ms.
          The extrapolation to $\Delta T_{win} = 0$ is shown as a dotted line.
          Error bars represent the uncertainties of the fit to the
          2.2\,MeV peak in the gamma spectra.}
  \label{fig:numbers} 
\end{figure}

Figure~\ref{fig:numbers} shows the time evolution of the signal, 
${N_{peak}}^{inside}$, accumulated in the 2.2\,MeV peak in the gamma spectrum
with growing $\Delta T_{win}$ on a linear time scale.
The values of ${N_{peak}}^{inside}$ 
are correlated, since any $\Delta T_{win}$ window contains all the events
already recorded in the smaller time windows.
The linear fit to the histogram for $\Delta T_{win} \ge 2$\,ms confirms
a flat background rate. The intercept with the $y$-axis provides the
total number of 2.2\,MeV gammas that can be attributed to the signal.

\begin{figure}[!h]
  \begin{center}
    \includegraphics[width=0.8\textwidth]
    {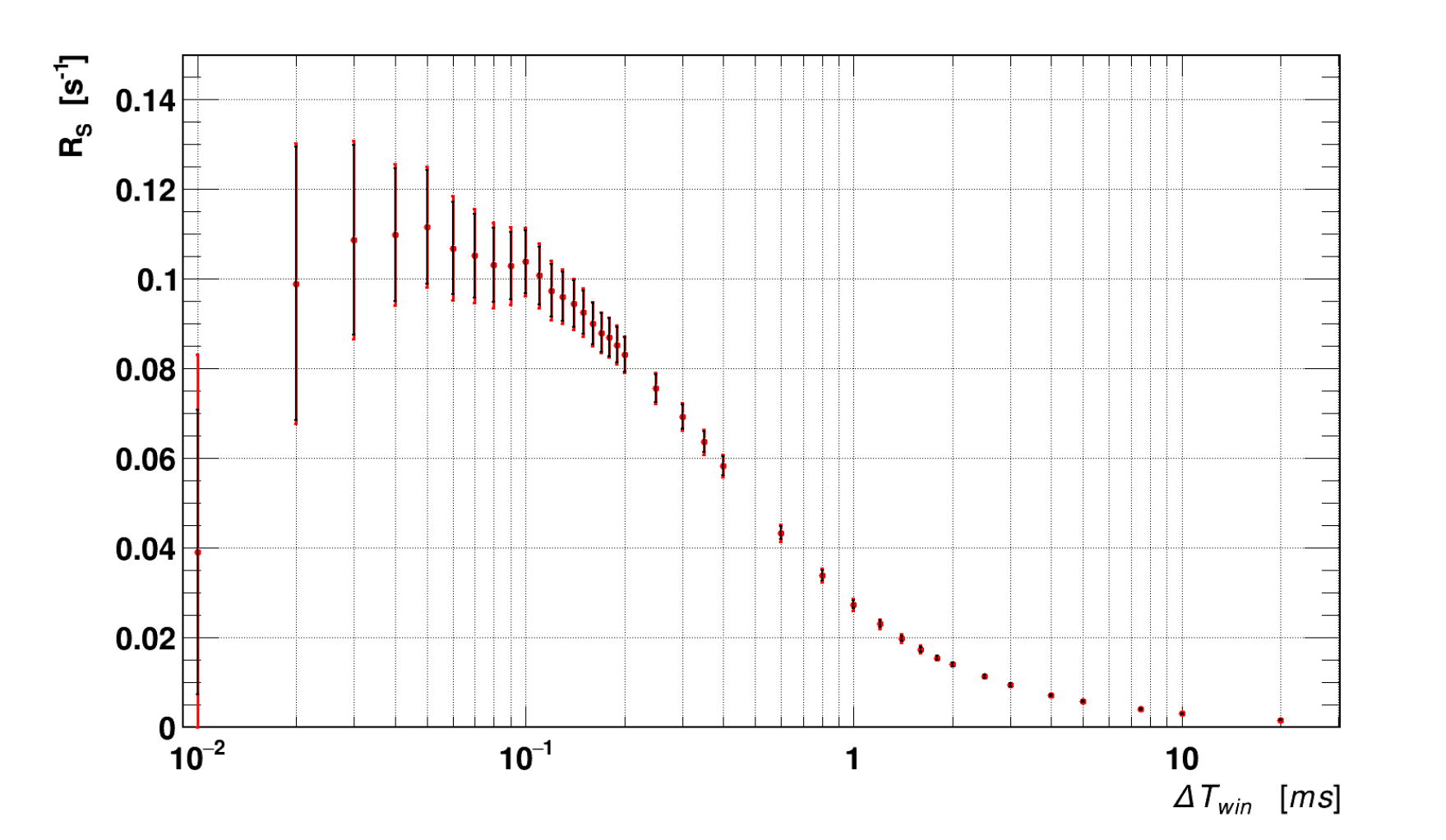}
  \end{center}
 \caption{Effective signal rate, ${R_S}$, vs.\ $\Delta T_{win}$.
  The inner error bars represent the statistical uncertainties, the
  outer bars represent the statistical and the systematic uncertainties 
  added in quadrature. Adapted from \cite{palermo}.}
  \label{fig:S-rate-data-fit-with-systematic-error} 
\end{figure}

\begin{figure} [!h]
  \begin{center}
    \includegraphics[width=0.8\textwidth]
   {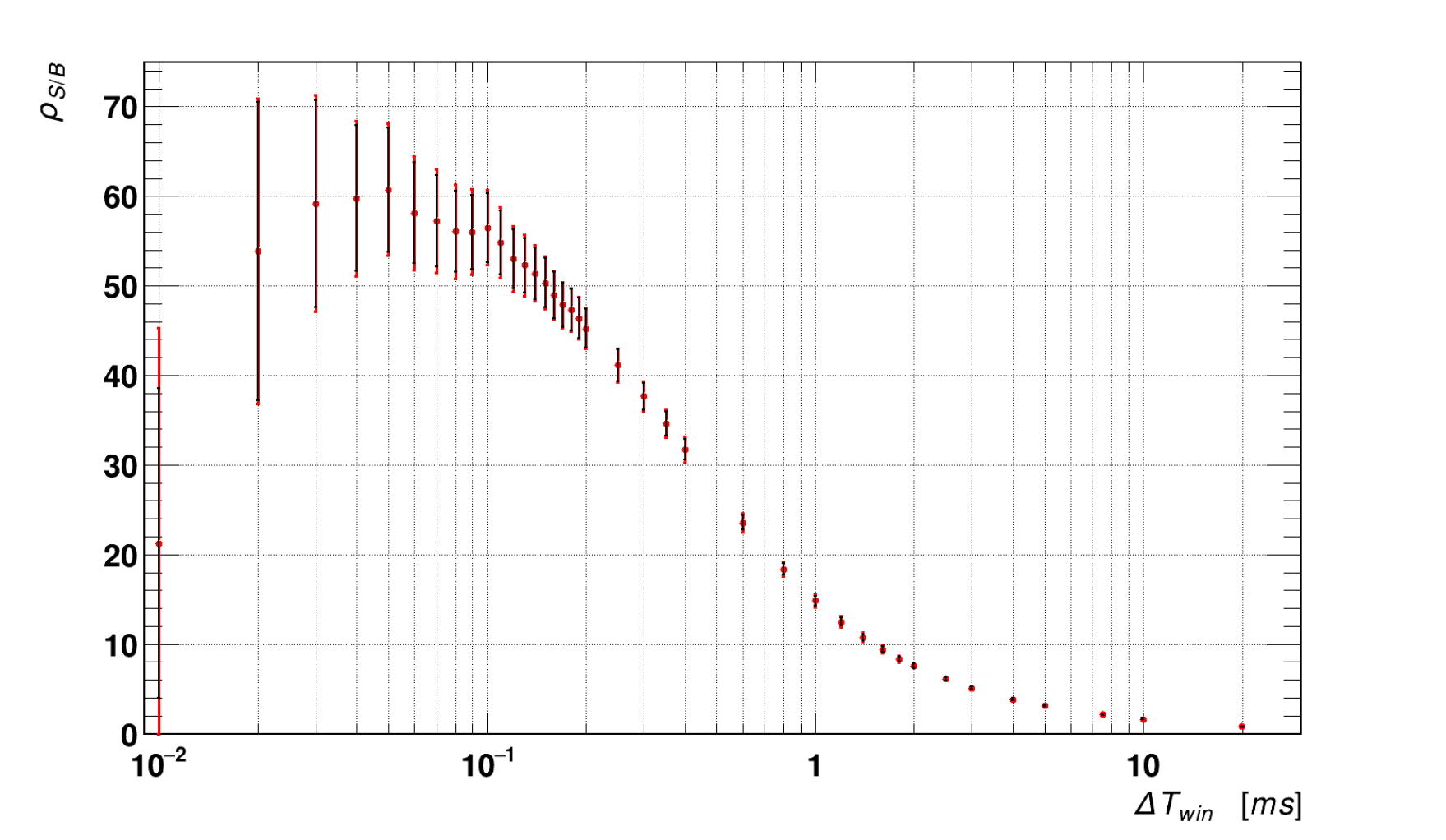}
  \end{center}
 \caption{Signal-to-background ratio, ${\rho_{S/B}}$, vs.\ $\Delta T_{win}$.  
  The inner error bars represent the statistical uncertainties, the
  outer bars represent the statistical and the systematic uncertainties 
  added in quadrature. Adapted from \cite{palermo}.}
  \label{fig:SB-ratio-data-fit-with-systematic-error} 
\end{figure}

The time evolution of the background-subtracted signal rate, $R_S$, is
shown in Fig.\,\ref{fig:S-rate-data-fit-with-systematic-error} 
on the exponential time scale also used for the visualisation
of the background rate
in Fig.\,\ref{fig:B-rate-data-fit-with-systematic-error}.  
The effective rates are the highest for $\Delta T_{win}$ between 20\,$\mu$s
and 100\,$\mu$s. The rates then drop,
which is equivalent to a decreasing number of
signal events added per unit time.
Beyond 1\,ms, no new signal events are recorded and thus the 
effective rate drops 
due to the larger windows considered.
Clearly, $\Delta T_{win}$ up to 1\,ms can be chosen 
for all studies of the signal. 
The signal-to-background ratios, $\rho_{S/B} =  {R_S} / {R_B }$,
for such windows are shown to be above 15 in 
Fig.~\ref{fig:SB-ratio-data-fit-with-systematic-error}. 
For $\Delta T_{win}$ between 20\,$\mu$s
and 100\,$\mu$s,  $\rho_{S/B}$ reaches values of up to 60. 

\newpage
\section{Comparison of Monte Carlo to data}
\label{sec:data-mc-comp}

The run\,I trigger configuration does not facilitate to distinguish between
events in the side-wall and elsewhere in the apparatus.
The resulting definition of signal and background for run\,I 
makes a direct physics interpretation of the data in terms of
the neutron production rate in lead difficult.
However, the data can be compared to Monte Carlo predictions.

\begin{figure} [!h]
 \centering
 \includegraphics[width=0.8\textwidth]
  {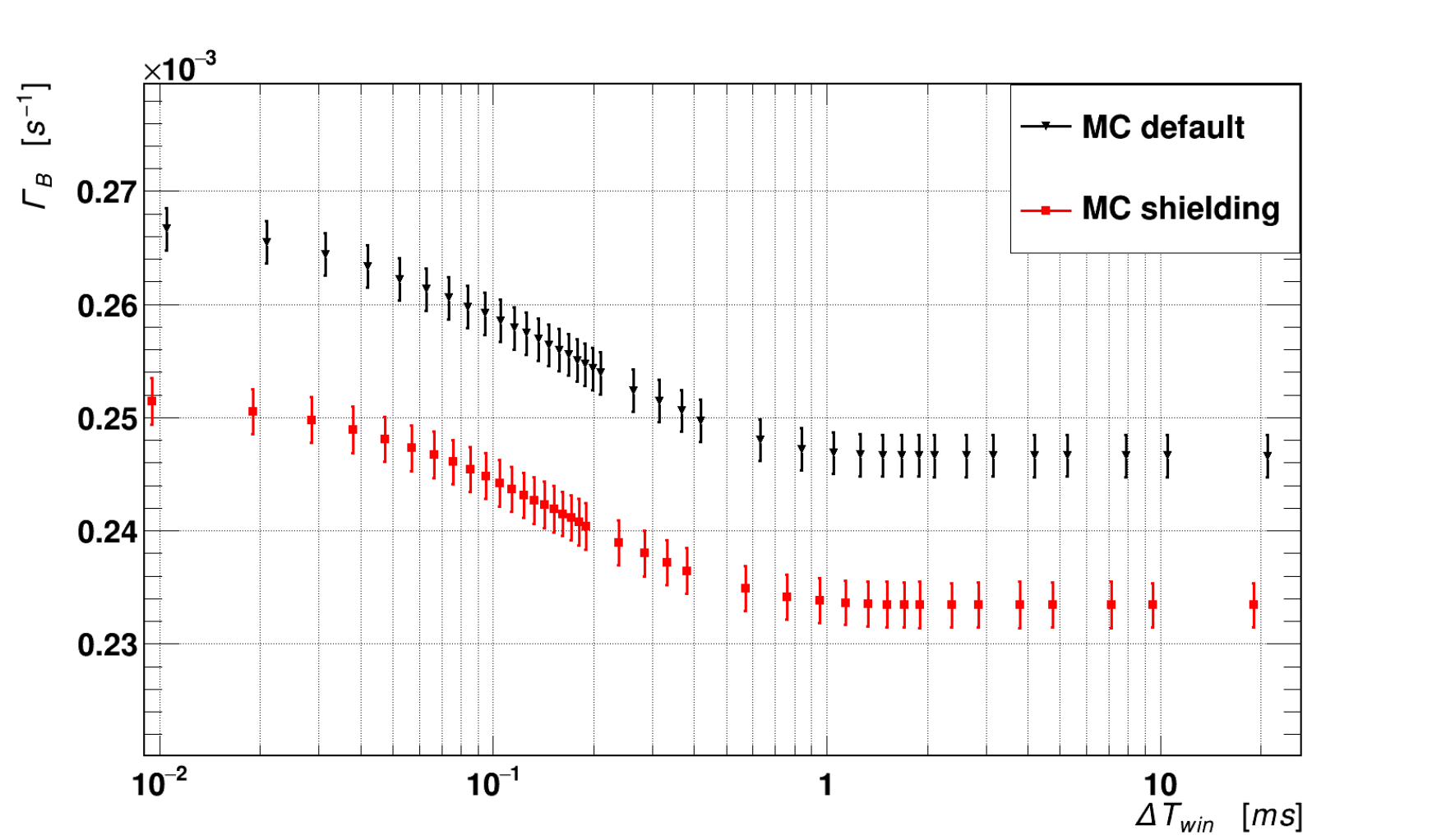}
\caption{Effective MC rate, $\Gamma_B^{MC}$, see Eq.~\ref{eqn:rate-inside},
         vs.\ $\Delta T_{win}$
         for the physics lists default and shielding.
         } 
\label{fig:data-MC-B-rate-before-correction-zeroMono}
\end{figure}

The simulations were performed within the Monte Carlo (MC)
framework MaGe~\cite{mage}, jointly developed by the GERDA~\cite{gerda} 
and MAJORANA~\cite{majorana} collaborations. 
Two different ``physics lists'' were used:
\begin{itemize}
 \item \textbf{default} as recommended by the MaGe developers;
 \item \textbf{shielding} as recommended by the
                   GEANT4 collaboration \cite{g4recommendation}.
\end{itemize}
These two physics lists differ mainly on the production of
neutrons in the primary interaction of the muon and in secondary
hadronic interactions. They are basically identical on neutron transport.

A simplified version~\cite{palermo} of the overburden of the TSUL, 
together with a detailed description of the MINIDEX run\,I apparatus 
were implemented in GEANT4 (version\,9.6.4~\cite{g4release}).
The cosmic muon~\cite{muspec} and neutron~\cite{nspec} spectra  
were used as input; both, muons and neutrons were tracked 
into the experimental volume, where their interactions were simulated.
  
The energy resolution of the HPGe detectors was taken into account 
by smearing all energies according to a Gaussian distribution 
with a constant standard deviation of 1.6\,keV. This value corresponds 
to a FWHM of $\approx 2.8$\,keV as measured for the peak at 2.2\,MeV. 
Since all the comparisons of MC to data 
only refer to the region around 2.2\,MeV, the simplification of a
constant resolution is justified.

The background components due to cosmic muons and neutrons were
simulated together with the signal. 
The background contribution from cosmic neutrons was found to be
around one order of magnitude smaller than from non-triggering
cosmic muons~\cite{palermo}. 

\begin{figure}[!h]
  \begin{center}
    \includegraphics[width=1\textwidth]
     {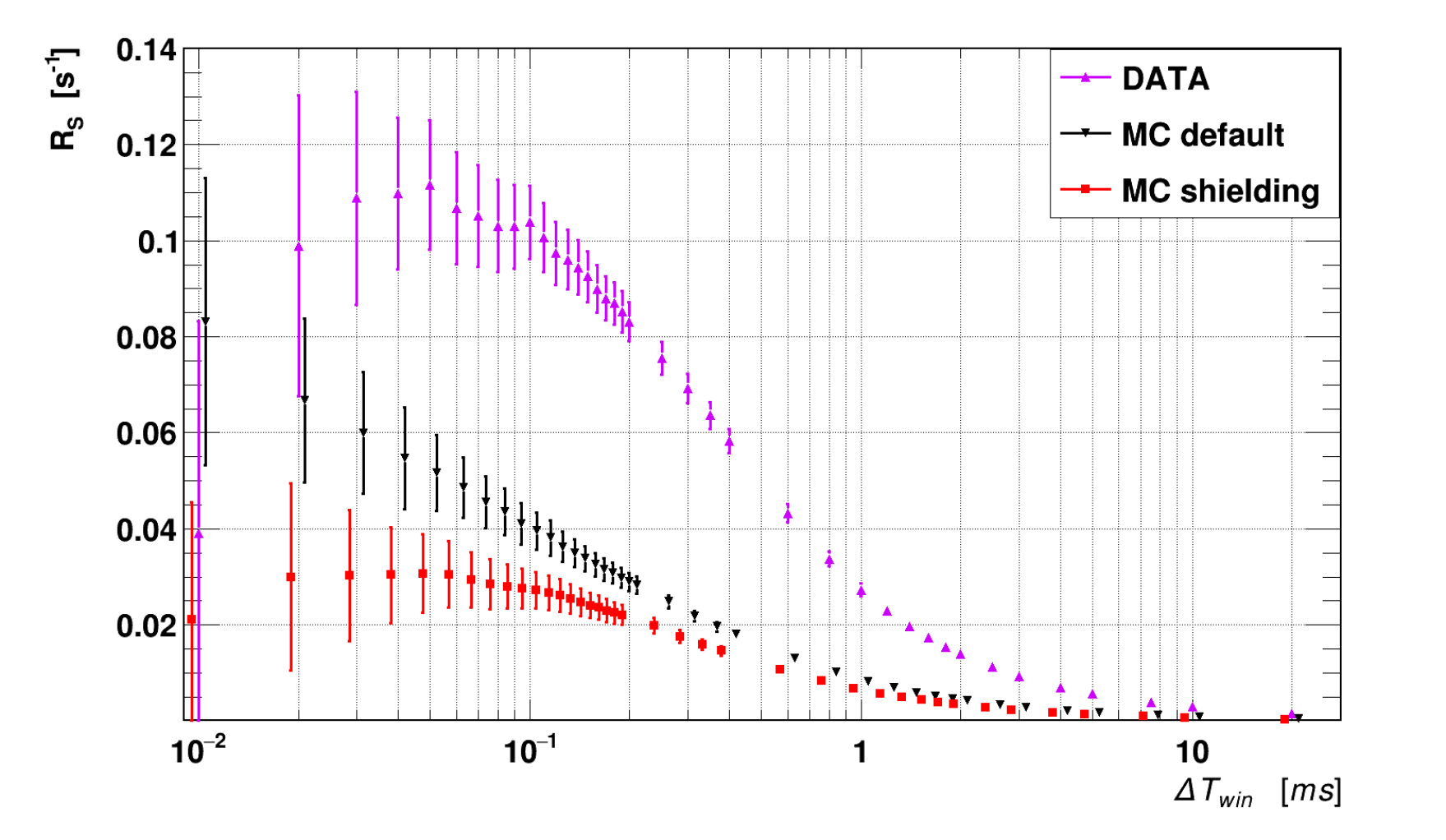}
  \end{center}
  \caption{Comparison of the MC predictions 
           from the two physics lists for $R_S$ to data.
           The points are horizontally displaced for better visibility.
           The error bars represent the total uncertainties,
           which are highly correlated.
           }
  \label{fig:data-MC-S-rate-before-correction-zeroMono} 
\end{figure}

The analysis of the MC data was performed exactly as for data
with equivalent definitions for $\Gamma_B^{MC}, R_S^{MC}$.
Figure\,\ref{fig:data-MC-B-rate-before-correction-zeroMono}
shows the time evolution of the  
rate $\Gamma_B^{MC}$ vs. $\Delta T_{win}$.
The background contribution has already flattened out 
at 1\,ms where the influence of the signal ends.
This is slightly earlier than observed for the data, see
Fig.~\ref{fig:B-rate-data-fit-with-systematic-error}. 
The prediction for $\Gamma_B^{MC}$ from the default list is 
about 5\,\% higher than from the shielding list 
for $\Delta T_{win} > 1$\,ms.

The component due to natural radioactivity was not included in the
simulation. This creates a large discrepancy in the 
absolute background rate between the MC and data. 
The background measured with $\Delta T_{win} = 4$\,ms
is more than a factor of seven higher in the data than predicted
for the cosmogenic contribution alone. This shows, that the background due
to natural radioactivity is dominating in the data.

The background-subtracted signal rate $R_S$, see Eq.\,\ref{eq:Rs}, 
allows a direct comparison of Monte Carlo
predictions to data. This is shown in
Fig.~\ref{fig:data-MC-S-rate-before-correction-zeroMono}.
Both physics lists predict significantly smaller signals than observed.
Table~\ref{tab:data-mc-S-real-rate-comparison-before-correction} lists 
${R_S}$  for $\Delta T_{win} = 4$\,ms as measured and predicted 
by the two MC versions. 
The data exceed the predictions from the MaGe default 
(GEANT shielding) Monte Carlo by a factor of  3.3 (4.1).
The MaGe default list predicts a strong prompt signal, not observed in
the data while the GEANT shielding list qualitatively describes 
the time evolution reasonably well.

\begin{table}[!h]
\renewcommand*{\arraystretch}{1.4}
\begin{center}
  \begin{tabular}{l|c}
    \hline
    \hline
            \multicolumn{2}{c}{Effective Signal Rate $R_S$  for} \\
    \hline
       Data                         &  (7.0 $\pm$ 0.2) $10^{-3}$\,Hz  \\   
     \hline
       GEANT with ``MaGe default'' list   &   (2.1 $\pm$ 0.1)  $10^{-3}$\,Hz \\
     \hline
       GEANT with ``GEANT shielding'' list  &  (1.7 $\pm$ 0.1) $10^{-3}$\,Hz \\ 
     \hline
  \end{tabular}
 \end{center}
 \caption{Effective signal rate, $R_S$, 
          for $\Delta T_{win}$\,=\,4\,ms for data and the two MCs.}
   \label{tab:data-mc-S-real-rate-comparison-before-correction} 
\end{table}

Another approach is to compare the MC to data before background
subtraction.
The muon-induced components of the MC predictions are separated
out by fitting the MC predictions to the data for each  $\Delta T_{win}$:
\begin{eqnarray}
(\Gamma_{S+B})_i ~= & A_i & \cdot ~~[~(\Gamma_{tr-\mu}^{inside})_i^{MC} 
                  + \Gamma_{ntr-\mu}^{MC}~] 
                  ~+~ B_i \cdot \Gamma_{neutrons}^{MC} \\ 
            R_B ~= & A_i & \cdot ~~\Gamma_{ntr-\mu}^{MC} 
                  ~+~ B_i \cdot \Gamma_{neutrons}^{MC}
\label{eqn:sim-correction1}
\end{eqnarray}
where:
\begin{itemize}
  \item the index $i$ is the ascending index for the different $\Delta T_{win}$;
  \item $A_i$ and $B_i$ are the fit parameters for window $i$.
        $A_i$ represent the factors by which the muon-induced components 
        are scaled to fit the data. 
        $B_i$ represent the factors by which the background 
        due to cosmic neutrons is scaled to account for the non-simulated
        background due to natural radioactivity;
  \item $\left({\Gamma_{S+B}}\right)_i$ is the measured $\Gamma_{S+B}$ 
        for window $i$;
  \item ${R_B}$ is the measured background rate;
  \item $(\Gamma_{tr-\mu}^{inside})_i^{MC} $ is the simulated effective 
        signal rate due to triggered muons for window $i$;
  \item $\Gamma_{ntr-\mu}^{MC}$ 
        is the simulated background rate due to non-triggered muons; 
  \item $\Gamma_{neutrons}^{MC}$ 
        is the simulated background rate due to 
        cosmic neutrons.
\end{itemize}

\begin{figure}[!h]
  \begin{center}
    \includegraphics[width=1\textwidth]
    {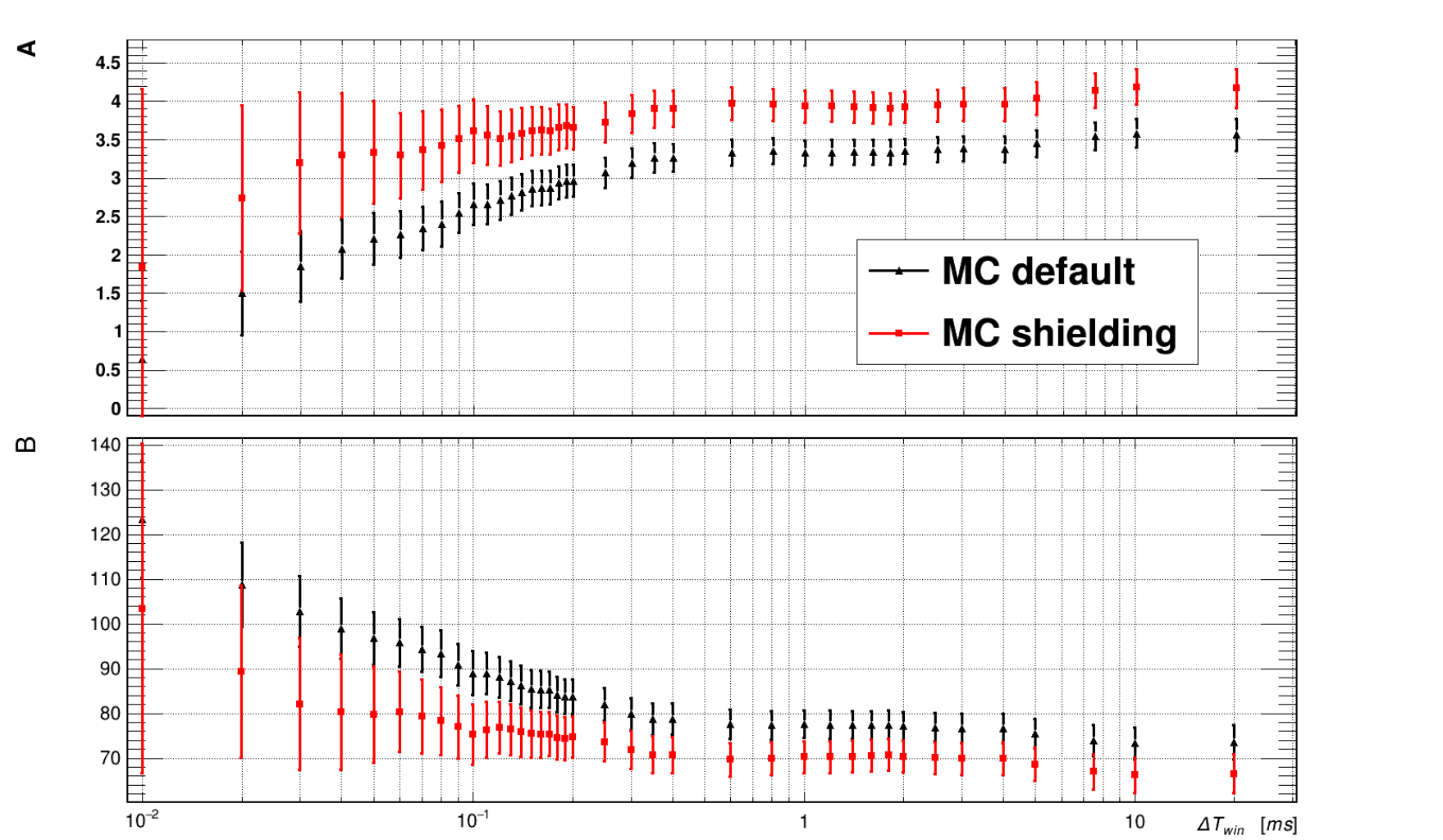}
  \end{center}
  \caption{Scale parameters $A_i$ and $B_i$ vs.\ $\Delta T_{win}$ as 
           obtained for the two different physics lists. 
           The error bars represent statistical uncertainties only.
           They are highly correlated as each time window contains
           all events recorded already in smaller time windows.}
  \label{fig:correction-factors-comparison} 
\end{figure}

It is possible to scale up the background due to cosmic neutrons
to include the background due to natural radioactivity,
because both background contributions are
constant in time and only the 2.2\,MeV gamma peak is considered. 
The results of the fits are depicted 
in Fig.~\ref{fig:correction-factors-comparison} and listed in
Table\,\ref{tab:correction-factors}.

If the time evolution of the signal was perfectly described by the 
simulation, the parameters $A_i$ and $B_i$ would be constant for all $i$.
However, as already seen 
in Fig.~\ref{fig:data-MC-S-rate-before-correction-zeroMono}, this is
not the case.
The values of $B_i$ are actually quite stable over most of the range.
At very small $\Delta T_{win}$, where the statistical uncertainty is large,
the $B_i$ increase as the procedure tries to compensate for the lacking 
signal strengths in the MCs.
As the shielding list describes the time evolution of the signal
reasonably well, the parameters $A_i$ vary not so much.
For the MaGe default physics list, the value of $A_1$ actually drops
below one. This confirms that this MC predicts a fast
component of the signal not present in the data.

\begin{table}[!h]
\renewcommand*{\arraystretch}{1.2}
\begin{scriptsize}
\begin{center}
  \begin{tabular}{c|c|c|c|c}
    \hline
    \hline
                  & \multicolumn{2}{c|}{$A_i$} & \multicolumn{2}{c }{$B_i$}\\
    $\Delta T_{win}$ & MaGe    & GEANT      & MaGe    & GEANT             \\
       $[$ms$]$         & default & shielding  & default & shielding      \\
    \hline
     0.01& 
     0.64 $\pm$ 0.77  &  1.84 $\pm$ 2.33& 123 $\pm$ 13 & 104 $\pm$ 37 \\
    \hline
     0.02& 
     1.50 $\pm$ 0.55  & 2.74 $\pm$ 1.21 & 109 $\pm$ 9  &  89 $\pm$ 19 \\
    \hline
     0.03& 
     1.85 $\pm$ 0.46  & 3.19 $\pm$ 0.92 & 103 $\pm$ 8  &  82 $\pm$ 15 \\
    \hline
     0.04& 
     2.08 $\pm$ 0.39  & 3.30 $\pm$ 0.81 &  99 $\pm$ 7  &  80 $\pm$ 13 \\
    \hline
     0.05& 
     2.20 $\pm$ 0.33  & 3.34 $\pm$ 0.67 &  97 $\pm$ 6  & 80 $\pm$ 11   \\   
    \hline
     0.06&  
     2.27 $\pm$ 0.30  & 3.30 $\pm$ 0.56 &  96 $\pm$ 5  & 80 $\pm$ 9    \\   
    \hline
     0.07& 
     2.35 $\pm$ 0.28  & 3.36 $\pm$ 0.51 &  94 $\pm$ 5  & 79 $\pm$ 8     \\   
    \hline
     0.08& 
     2.40 $\pm$ 0.29  & 3.43 $\pm$ 0.47 &  93 $\pm$ 5  & 78 $\pm$ 8    \\   
    \hline
     0.09&  
     2.55 $\pm$ 0.26  & 3.51 $\pm$ 0.43 &   91 $\pm$ 5 & 77 $\pm$ 7     \\   
    \hline
     0.10& 
     2.66 $\pm$ 0.27  & 3.61 $\pm$ 0.42 & 89 $\pm$ 5   & 75 $\pm$ 7    \\   
    \hline
     0.11&   
     2.66 $\pm$ 0.26  & 3.55 $\pm$ 0.38 & 89 $\pm$ 5   & 76 $\pm$ 6    \\   
    \hline
     0.12&  
     2.71 $\pm$ 0.25  & 3.52 $\pm$ 0.36 & 88 $\pm$ 5   & 77 $\pm$ 6    \\   
    \hline
     0.13&  
     2.77 $\pm$ 0.24  & 3.55 $\pm$ 0.34 & 87 $\pm$ 4   & 76 $\pm$ 6     \\   
    \hline
     0.14&  
     2.82 $\pm$ 0.24  & 3.58 $\pm$ 0.33 & 86 $\pm$ 4   & 76 $\pm$ 5     \\   
    \hline
     0.15&  
     2.87 $\pm$ 0.23      &  3.61 $\pm$ 0.32  &    86 $\pm$ 4   &    75 $\pm$ 5        \\   
    \hline
     0.16&   2.87 $\pm$ 0.23    &  3.62 $\pm$ 0.31   &     85 $\pm$ 4  &    75 $\pm$ 5      \\   
    \hline
     0.17&  2.88 $\pm$ 0.22     &  3.61 $\pm$ 0.30    &     85 $\pm$ 4  &   75 $\pm$ 5     \\   
     \hline
     0.18&   2.94 $\pm$ 0.22    &  3.66 $\pm$ 0.30   &     84 $\pm$ 4  &  75 $\pm$ 5     \\   
     \hline
     0.19&   2.97 $\pm$ 0.21    &  3.68 $\pm$ 0.29  &     84 $\pm$ 4  &  74 $\pm$ 5      \\   
     \hline
     0.20&   2.96 $\pm$ 0.21    &  3.65 $\pm$ 0.28    &     84 $\pm$ 4  &  75 $\pm$ 5     \\   
     \hline
     0.25&  3.07 $\pm$ 0.20     &   3.72 $\pm$ 0.26    &     82 $\pm$ 4  &  74 $\pm$ 4    \\   
     \hline
     0.30&   3.20 $\pm$ 0.19    &  3.83 $\pm$ 0.25     &     80 $\pm$ 4  &  72 $\pm$ 4   \\   
     \hline
     0.35&   3.26 $\pm$ 0.19    &  3.90 $\pm$ 0.24   &      79 $\pm$ 4 &   71 $\pm$ 4    \\   
     \hline
     0.40&    3.26 $\pm$ 0.18   &   3.91 $\pm$ 0.23  &       79 $\pm$ 3&   71 $\pm$ 4     \\   
     \hline
     0.60&   3.33 $\pm$ 0.17    &  3.97 $\pm$ 0.22 &      78 $\pm$ 3 &   70 $\pm$ 4        \\   
     \hline
     0.80&   3.34 $\pm$ 0.17    &  3.96 $\pm$ 0.21 &       77 $\pm$ 3&   70 $\pm$ 4    \\   
     \hline
     1.00&   3.33 $\pm$ 0.16    &  3.94 $\pm$ 0.21 &        78 $\pm$ 3 &     70 $\pm$ 4    \\   
     \hline
     1.20&   3.34 $\pm$ 0.16    &  3.94 $\pm$ 0.21  &       78 $\pm$ 3&   70 $\pm$ 4      \\   
     \hline
     1.40&    3.34 $\pm$ 0.16   &  3.93 $\pm$ 0.21   &      77 $\pm$ 3 &   70 $\pm$ 4    \\   
     \hline
     1.60&   3.34 $\pm$ 0.16    &   3.91 $\pm$ 0.20  &      77 $\pm$ 3 &   71 $\pm$ 4     \\   
     \hline
     1.80&    3.34 $\pm$ 0.16   &   3.91 $\pm$ 0.20  &     78 $\pm$ 3  &    71 $\pm$ 4   \\   
     \hline
     2.00&    3.35 $\pm$ 0.16   &   3.93 $\pm$ 0.21  &    77 $\pm$ 3  &    70 $\pm$ 4    \\   
     \hline
     2.50&   3.37 $\pm$ 0.17    &   3.95 $\pm$ 0.21  &   77 $\pm$ 3   &     70 $\pm$ 4    \\   
     \hline
     3.00&   3.38 $\pm$ 0.17    &   3.96 $\pm$ 0.21   &    77 $\pm$ 3   &    70 $\pm$ 4   \\   
     \hline
     4.00&   3.38 $\pm$ 0.17    &   3.96 $\pm$ 0.21  &    77 $\pm$ 3   &    70 $\pm$ 4    \\   
     \hline
     5.00&   3.45 $\pm$ 0.17    &   4.04 $\pm$ 0.21  &     76 $\pm$ 3  &    69 $\pm$ 4   \\   
     \hline
     7.50&   3.54 $\pm$ 0.18    &   4.14 $\pm$ 0.23   &      74 $\pm$ 3 &   67 $\pm$ 4    \\   
     \hline
     10.00&    3.58 $\pm$ 0.19    &   4.19 $\pm$ 0.23  &      73 $\pm$ 4 &   66 $\pm$ 4   \\   
     \hline
     20.00&   3.56 $\pm$ 0.21    &    4.17 $\pm$ 0.25  &    74 $\pm$ 4   &   67 $\pm$ 4      \\        
    \hline
    \hline
 \end{tabular}
\end{center}
\end{scriptsize}
\caption{Values of the scale factors $A_i$ and $B_i$ for each $\Delta T_{win}$
        for both MC predictions. Uncertainties are statistical only.} 
\label{tab:correction-factors} 
\end{table}

\section{Summary and Outlook}
\label{out}

A new experiment to measure the production of neutrons by cosmic muons,
MINIDEX, was introduced. It is a compact apparatus, easy to move
to different locations, almost maintenance free and operated remotely.
Neutrons are detected after thermalisation through 2.2\,MeV gammas emitted
after their capture by the hydrogen in water.
Thus, MINIDEX measures neutrons without an energy threshold. 
The signal and the background are measured simultaneously, allowing
the extraction of the signal and its time evolution
without major assumptions or Monte Carlo simulations.

The results of the first run of MINIDEX with lead as the target
material and at a depth of about 16\,meter water equivalent were presented. 
A clear signal was observed with signal-to-background ratios 
above 15 for the complete development of the
signal over one millisecond.
The results were compared to GEANT\,4 based Monte Carlo
simulations using the so called 
``MaGe default'' and ``GEANT shielding'' 
physics lists. Both lists led to predictions that underestimate
the integrated signal strength significantly.  
The data show an overall production of  
muon-induced 2.2\,MeV gammas a factor of 3.3 (4.1) higher
than predicted using the MaGe default (GEANT\,4 shielding) list.
While the shielding list does not describe the overall signal strength  
as well as the MaGe default list
it qualitatively describes the time evolution
better. In particular, the MaGe default list predicts a 
prompt signal component not observed in the data.

During its first data taking period, 
MINIDEX ran with a very simple muon trigger.
In January 2016, the setup was upgraded to allow for the identification 
of different event topologies. Especially, muons can be identified
that pass vertically and only through the selected high-Z material.  
Different event topologies should also provide some handle
to disentangle the primary neutron production rate from
the influence of neutron transport.

At the end of 2016, a switch to copper as the target material is foreseen.
The data is expected to be available as input to 
the design phase of the next-generation large-scale
low-background experiments currently envisioned, where lead and copper
are discussed as shielding materials.
During the following years,
MINIDEX will explore the neutron production 
due to cosmic muons in a variety of high-Z
materials. In addition, the dependence on the
muon energy can be probed by moving MINIDEX to different depths.

\section{Acknowledgments}
\label{ack}
We would like to thank the technical department of the 
Max-Planck-Institut f\"{u}r Physik for their strong support.
We would also like to thank the Universit\"at T\"ubingen for the space in TSUL, especially Peter Grabmaier and Igor Usherov for their 
hospitality and help.

\vskip 1 cm
{\noindent}
{\bf Bibliography:}

\bibliographystyle{elsarticle-num}
\bibliography{<your-bib-database>}

\end{document}